\documentclass[prl,superscriptaddress,notitlepage,showpacs,longbibliography,twocolumn]{revtex4-2}
\usepackage{float,subfig,xcolor,latexsym,mathrsfs}
\usepackage{amsmath}
\usepackage{amssymb}
\usepackage{graphicx,array}
\usepackage{caption}
\usepackage{txfonts}
\usepackage{mathtools}

\newcommand{\beq}{\begin{equation}}
\newcommand{\eeq}{\end{equation}}
\graphicspath{{Figstore/}}

\begin{document}

\title{Full quantum theory of nonequilibrium phonon condensation and phase transition}

\author{Xuanhua Wang}
\email{wangxh@ucas.ac.cn}
\affiliation{Center for Theoretical Interdisciplinary Sciences, Wenzhou Institute, University of Chinese Academy of Sciences, Wenzhou, Zhejiang 325001, China}

\author{Jin Wang}
\email{jin.wang.1@stonybrook.edu}
\affiliation{Center for Theoretical Interdisciplinary Sciences, Wenzhou Institute, University of Chinese Academy of Sciences, Wenzhou, Zhejiang 325001, China}
\affiliation{Department of Physics and Astronomy, Stony Brook University, Stony Brook, New York 11794, USA}
\affiliation{Department of Chemistry, Stony Brook University, Stony Brook, New York 11794, USA}

%\date{\today}

\begin{abstract}
Fr\"olich condensation is a room-temperature nonequilibrium phenomenon which is expected to occur in many physical and biological systems. Though predicted theoretically a half century ago, the nature of such condensation remains elusive. In this Letter, we derive a full quantum theory of Fr\"ohlich condensation from the Wu-Austin Hamiltonian and present for the first time an analytical proof that a second-order phase transition induced by nonequilibrium and nonlinearity emerges in the large-$D$ limit with and without decorrelation approximation. This critical behavior cannot be witnessed if external sources are treated classically. We show that the phase transition is accompanied by large fluctuations in the statistical distribution of condensate phonons and that the Mandel-Q factor which characterizes fluctuations becomes negative in the limit of excessive external energy input. In contrast with the cold atom equilibrium BEC, the Fr\"ohlich condensate is a result of the nonequilibrium driving where the pump plays a role of setting the number of particles, and the medium plays a role of setting the temperature. Hence, BEC can either arise by reducing the medium temperature at fixed pump (equilibrium case), or by increasing the pump at fixed medium temperature (nonequilibrium case).

%Bose-Einstein condensation (BEC) was predicted a century ago, though its existence was expected and invoked little doubt, its discovery was only made possible in the last 30 years due to the technological advancement of the cooling method. On the other hand,   This leads to divided opinions on the nature of the process.
%we study the the Fr\"ohlich condensate and report for the first time the qualitative behavior as well as the analytical results of the nonequilibrium-induced phase transition at room temperatures using a full quantum approach. %Such transitions can be realized in out-of-equilibrium quantum systems such as Fr\"ohlich condensate and phonon-polaritons. 
%Note on the Fr\"olich condensate driven by two baths. We find that the nonequilibrium condensate of phonons can be induced by imposing two different baths on a system. 
\end{abstract}
\maketitle
\thispagestyle{empty}

%\noindent Please note: Abbreviations should be introduced at the first mention in the main text – no abbreviations lists. Suggested structure of main text (not enforced) is provided below.

\textit{Introduction.}-- In biological systems, a driven nonequilibrium condensation phenomenon which is often compared with the Bose-Einstein condensation (BEC) was hypothesized by Fr\"olich in the late 1960s \cite{frohlich1968long,frohlich1968bose,frohlich1970long,Frohlich1988}. While BEC is a low-temperature equilibrium phenomenon whose discovery was only recently realized as the cooling technology advances \cite{anderson1995observation,davis1995bose}, Fr\"ohlich condensate is intrinsically an out-of-equilibrium condensation of collective vibrations at the lowest mode driven by an external energy supply and is expected to occur at much higher temperatures \cite{bunkov2013spin}. The proposal has stimulated a wide range of studies on the collective modes in nonequilibrium systems such as quasi-equilibrium magnon condensation at room temperature \cite{demokritov2006bose,chumak2009bose}, polariton condensation in photonic systems \cite{kasprzak2006bose,manni2011polariton,deng2010exciton,zhang2022quantum,bloch2022non}, as well as nonequilibrium phase transitions \cite{haken1975cooperative,haken1977synergetics}. Recently, due to the rapid progress of terahertz technology, many experimental and analytical works have been carried out in this direction and signals of the long-anticipated nonequilibrium condensation were finally observed \cite{pokorny2004excitation,reimers2009weak,turton2014terahertz,lundholm2015terahertz,nardecchia2018out}. Further experimental investigations in both biological and non-biological systems have been proposed, such as spectroscopic studies using lysozymes and the bovine serum albumin (BSA) proteins, and optomechanical experiments using an array of membranes coupled with an cavity \cite{zhang2019quantum,zheng2021frohlich}. 

The essential piece in Fr\"ohlich's conjecture is the rate equations of the vibrational modes. The rate equations in Fr\"ohlich's reasoning were constructed heuristically by requiring a Bose-Einstein distribution of the phonons in the absence of energy sources \cite{frohlich1968bose,frohlich1968long}. A rigorous derivation of the rate equations was provided by T.M.~Wu and S.~Austin from a quantum Hamiltonian, now known as Wu-Austin Hamiltonian \cite{wu1977bose,wu1978bose,wu1978cooperative,wu1981frohlich}. The Hamiltonian is composed of three parts: the oscillating dipoles, the heat bath that interacts nonlinearly with the system dipoles and the external energy supply. The microscopic theory was formulated based on the Hamiltonian and finite-temperature Green's functions, and confirmed the possibility of detecting the phonon condensation in biological systems even at room temperature \cite{wu1977bose,wu1978bose,wu1978cooperative,wu1981frohlich}. The rate equations suggest the emergence of the cooperative self-organization when the input energy is strong and the chemical potential approaches the lowest energy in the branch of vibrational modes \cite{wu1978cooperative,preto2017semi}.

Though the theory of Fr\"ohlich condensation dates back half a century ago, little was known regarding its quantum and statistical properties near the transition. The common lore for predicting the feature of Fr\"ohlich condensate is to make an analogy of the atomic BEC \cite{frohlich1968long,frohlich1968bose,frohlich1970long,Frohlich1988}. However, this \textit{a priori} assumption should be carefully scrutinized as Fr\"ohlich condensation is intrinsically an out-of-equilibrium phenomenon whose mechanism is more suitably compared to that of a single mode maser, of which the quantum properties and phase transitions are intensively studied \cite{rice1994photon,scully1966quantum,scully1967quantum,scully1999condensation,degiorgio1970analogy,graham1970laserlight,zhang2019quantum,zhang2021quantum}. The first investigation of the quantum statistics of the condensate was done in Ref.~\cite{zhang2019quantum} where the system and the bath are modeled by quantum oscillators and the external pump is treated classically.  The calculation in \cite{zhang2019quantum} shows a continuous crossover between the two phases--the normal phase with almost no condensate and the condensate phase. Tough some critical pump strength can be defined artificially as an reference scale at which the external input is roughly strong enough to generate the condensate, there is no qualitative distinction across the two ``phases". On the other hand, recent experiments using BSA proteins have indicated not only the existence of the collective oscillations of the biomolecules but also a sign of a sharp transition when the control parameter exceeds certain critical threshold value \cite{nardecchia2018out,reimers2009weak}. Like many studies on Fr\"ohlich condensate, the analysis in Ref.~\cite{nardecchia2018out} adopts a semi-classical method \cite{jauslin2010dynamics,preto2017semi}. They proceeded by dequantizing the Wu-Ausin Hamiltonian to obtain the classical equations of motion. However, such treatment renders the analysis nearly intractable and difficult to retrieve further information beyond the dynamics of mean values \cite{jauslin2010dynamics}. 

In this Letter, we derive a full quantum theory of the nonequilibrium condensation from the Wu-Austin Hamiltonian. In contrast with Ref.~\cite{zhang2019quantum}, both the surrounding solvents and the external source are treated quantum mechanically in our work, which results in crucial differences in the critical behaviors near the condensation point from what was reported in \cite{zhang2019quantum} such as the emergence of nonequilibrium phase transitions. We provide an analytical proof that the transition from the nonequilibrium steady state (NESS) with no condensate to the condensate state is similar to a second-order phase transition in the large-D limit regardless of certain approximations used in the precedent research. Moreover, the finite-size effects of the system on the transition and statistics are investigated. Unlike BEC, the condensation process does not require extremely low temperature but a high-temperature energy input. We demonstrate that such transition is accompanied by large fluctuations in its statistics and that excessive energy input promotes the system to the regime with a negative Mandel-Q parameter.

\textit{Driven Dissipative System.}--The appearance of Fr\"ohlich condensate is believed to be a general consequence of the following 
three conditions: dissipation to the thermal bath, external energy source, and nonlinearity. For the terahertz vibrations inside protein molecules such as phonon modes in DNA or BSA proteins, the surrounding solvents function approximately as a thermal bath. The nonlinear coupling between the system vibration modes and the surrounding medium in the second quantization form gives rise to the Hamiltonian formulized by Wu and Austin \cite{wu1977bose,wu1978bose,wu1978cooperative,wu1981frohlich}. The system we consider is modeled by a collection of oscillators in a narrow bandwidth $\omega_i\in \mathcal{I}_{sys}$ which corresponds to $D$ normal modes of the oscillation. The vibrational modes are annihilated by the operator $a_i$. The normal modes interact with the surrounding biological system with the excitation energies $\bar{\Omega}_k$ and the associated annihilation operator $b_k$, as well as the external source with frequency $\Omega_i$ and annihilation operator $p_i$. The interaction Hamiltonian can be written as:
\begin{equation}
\begin{split}
V_{\text{int}}(t) =& \hbar\sum_i\sum_k f_{i,k} a_i^{\dagger}b_k e^{i(\omega_i-\bar{\Omega}_k)t}  \\
& + \hbar\sum_{i,j}\sum_k g_{ij,k}a_i^{\dagger}a_j b_k e^{i(\omega_i-\omega_j-\bar{\Omega}_k)t} \\
&+\hbar\sum_i\sum_k \lambda_{i,k} a_i^{\dagger}p_k e^{i(\omega_i-\Omega_k)t} + \text{h.c.} \,,
\end{split}
\label{Vt}
\end{equation}
where $f_{i,k}$ and $g_{ij,k}$ are the coupling constants between the molecular vibrations and the solvent, and $\lambda_{i,k}$ is the coupling between the external source field and the system.  $\omega_i$, $\bar{\Omega}_k$ and ${\Omega}_k$ denote the frequencies associated with the $i$-th vibrational mode of molecules, the solvent modes and of the energy source. The Hamiltonian has been criticized for the unbounded potential energy from below \cite{bolterauer1999elementary}. This does not concern us here as the Hamiltonian is only used as an effective approximation in the interested energy domain rather than a fundamental description. 

The rate equation of the condensate can be derived directly from the above interaction Hamiltonian. The details of the derivation can be found in the Supplementary Material. We use the standard techniques of the full quantum master equation with Born and Morkov approximations which returns formally the equations of motion for the whole system \cite{scully1999quantum,breuer2002theory,agarwal2012quantum}. We then reduce the equations of motion of the density operator to only the $l$-th state. After rearranging and contracting the repeated indices, we obtain the rate equation of phonons at mode $\omega_l$ in reminiscence of the heuristic equation given by Fr\"ohlich:
\begin{equation}
\begin{split}
\langle \dot{n}_l\rangle  =
\ \phi & \big[\bar{n}_{\omega_l} - \langle n_l\rangle\big]+\Lambda \big[{n}_{\omega_l} - \langle n_l\rangle\big]\\
+&\chi\left\{\sum_{j>l}\big[\bar{n}_{\omega_{jl}}\langle n_j-n_l \rangle +\langle n_jn_l+n_j\rangle\big]\right. \\
 &\ \ \left. \ + \sum_{j<l}\big[\bar{n}_{\omega_{lj}}\langle n_j-n_l\rangle - \langle n_j n_l+n_l\rangle\big]\right\} \,,
\end{split}
\label{nl}
\end{equation}
where $\phi=2\pi f_{\omega}^2{\cal D}(\omega),\ \Lambda=2\pi \lambda_{\omega}^{2}{\cal D}(\omega),\ \chi=2\pi g_{\omega}^2{\cal D}(\omega)$ are the rates of dissipation, energy input and energy redistribution, respectively. ${\cal D}(\omega)$ is the density of states of the surrounding environment. Since the bandwidth of the oscillations $ \mathcal{I}_{sys}$ is assumed to be narrow, the variation of the bath density and the couplings within the frequency range is slow, thus the rates can be well approximated to be the same for each mode. As such, $\mathcal{D}(\omega),\, f_\omega,\, \lambda_\omega,\, g_\omega$ can all be evaluated at a typical value of $\omega_0$, the lowest of the vibrational modes. The rate equation of the total number of phonons $\hat N=\sum_{s=0}^D a_s^{\dagger}a_s$ can be further calculated to be the following: $\langle\hat{\dot{N}}\rangle  =(D+1)(\phi\bar{n}_s+\Lambda n_{ex}) - (\phi+\Lambda)\langle N\rangle$. In the NESS, the solution of the total number of phonons is:
\begin{equation}
    \langle N\rangle\simeq \frac{(D+1)(\phi\bar{n}_s+\Lambda n_{ex})}{\phi+\Lambda}\,,
    \label{cn}
\end{equation} 
where $\bar{n}_s$ is the occupation of the solvent at frequency $\omega_0$, $n_{ex}$ is the occupation of the external source at $\omega_0$ and $D$ is the total number of modes. Notice that the nonlinear term does not contribute to the total phonon number and only contributes to the energy redistribution. %For generality, we can assume the existence of the nonlinear coupling for the external energy source $\chi^\prime$. 
At the steady state, the phonon number at mode $\omega_0$ can be formally expressed as
\begin{equation}
\langle n_0\rangle = \frac{\phi\bar{n}_s+\chi(\bar{n}_s+1)(\langle N\rangle-\langle n_0 \rangle)+\Lambda n_{ex}}{\phi+\Lambda-\chi(\langle N\rangle-\langle n_0 \rangle-D\bar{n}_s)}\,.
\label{n0sol}
\end{equation}
It is easy to check that by setting the nonlinear coupling $\chi=0$, the mean value of phonon number at mode $\omega_0$ becomes the weighted average of thermal distributions due to the solvent and the external source: $\langle n_0\rangle=\dfrac{\phi \bar{n}_s+\Lambda n_{ex}}{\phi+\Lambda}$. This is expected from the secular approximation used in the equations of motion which ignores the off-diagonal subdominant contribution \cite{scully1999quantum,breuer2002theory,agarwal2012quantum,wang2019nonequilibrium,wang2022effect}. To obtain Eq.~\eqref{n0sol}, we have assumed the decorrelation approximation between the $n_0$ and $n_l$, namely $\langle n_0 n_l\rangle\simeq \langle n_0\rangle\langle n_l\rangle$ and $\langle n_0^2 \rangle\simeq \langle n_0\rangle^2$. In the case of large molecule numbers, the total number of vibrational modes $D$ is large and the solution for the mean condensate number is approximately:
\begin{align}
    \langle n_0 \rangle \simeq \frac{\langle N\rangle-D \bar n_s}{2}+\frac{1}{2}\sqrt{(\langle N\rangle-D\bar n_s)^2+4(\bar n_s+1)\langle N\rangle}\,.
    \label{semi_solution}
\end{align}
It is easy to see that in the limit of large total number of modes $D$, the fraction of the condensate becomes $ \langle n_0 \rangle/\langle N\rangle\xrightarrow{D\rightarrow \infty} 0$ for $n_{ex}\leq \bar{n}_s$. When $n_{ex} > \bar{n}_s$, we have:
\begin{align}
    \langle n_0 \rangle /\langle N\rangle \simeq \frac{\Lambda}{\phi \bar{n}_s+\Lambda n_{ex}}(n_{ex}-\bar{n}_s)\,,
    \label{semi_condition}
\end{align}
which suggests a continuous phase transition at $n_{ex}=\bar{n}_s$ and gives the critical exponent $\beta=1$. The fraction approaches to one in the limit of large $n_{ex}$.

%Figure 1
 \begin{figure}
    \centering
    \includegraphics[width=0.47\textwidth]{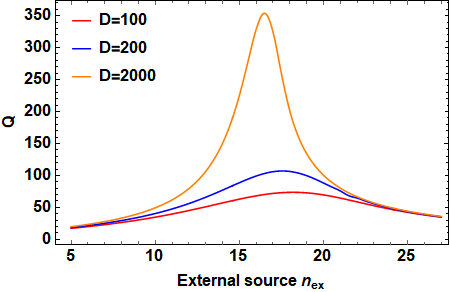}
    \includegraphics[width=0.47\textwidth]{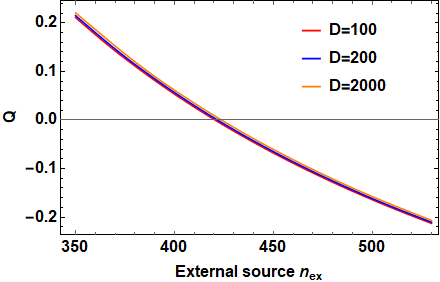}
    \caption{Mandel-$Q$ parameter for different values of the total mode number $D$. (a) The Mandel-$Q$ parameter reaches maximal near the phase transition points given by Eq.~\eqref{Sec2:condition}. (b) For the external source at high intensity, the Mandel parameter becomes negative. Parameters are $\bar{n}_s=16,\, \phi=5\mathrm{GHz}$, $\Lambda=10\mathrm{GHz}$, $\chi=0.07\mathrm{GHz}$. }
    \label{Fig1}
\end{figure}

\textit{Condensate statistics and fluctuations without decorrelation approximation.}--The above approach to solve for the approximate mean value of condensate phonons is a reminiscence of Fr\"ohlich's equation. However, one can argue that the decorrelation approximation used in obtaining the formula is heavily invalidated by the nonlinear term in the Hamiltonian, as well as by the coherence in the vibrational modes and fluctuations in each mode \cite{rice1994photon,zhang2021quantum}. Therefore, in situations with strong coherence or near the transition point, such approximations are not applicable and new approaches need to be considered. To gain insight into the transition phenomenon, we study the quantum statistics of condensate phonons assuming only the decorrelation between the total number of phonons $\langle N\rangle$ and the condensate phonon number $\langle n_0 \rangle$, namely, $\langle n_0 N\rangle\simeq \langle n_0\rangle\langle N\rangle$. The total phonon number $\langle N \rangle$ has a much smaller relative fluctuation due to the central limit theorem and is conserved under the nonlinear two-phonon interaction which correlates two individual modes. The statistics of the condensate contains not only the information about the mean value of the phonon numbers in the condensate but also its distribution. One of the important features of the distribution is captured by its the Mandel-Q parameter defined as $Q=\left(\langle n_0^2\rangle -\langle n_0 \rangle^2 \right)/\langle n_0 \rangle -1$, which characterizes the fluctuation of a distribution as well as its distance from the classical Poisson distribution \cite{mandel1982squeezed}. For any classical probability distribution the range of the Mandel parameter is $Q \ge 0$, where $Q=0$ corresponds to the Poisson distribution. Negative values of $Q$ means no classical analog and sub-Poissonian statistics exemplified by certain squeezed states of light \cite{davidovich1996sub}. 

We reduce the density matrix into the lowest diagonal mode to retrieve the information of only the condensate modes,  $\rho_{n_0,n_0} = \sum_{\{n_l\}}\langle n_0,\{n_l\}|\rho|n_0,\{n_l\}\rangle$, where $\{n_l\} = \{n_1,n_2,\cdots,n_D\}$ is the configuration of the excited states and $n_0$ is the variable denoting the phonon number in the condensate. For one specific $n_0$, the diagonal element has the physical interpretation of the probability of finding $n_0$ phonons on the lowest mode. The steady state solution for the distribution of the condensate phonons can be solved exactly. For details, please refer to Supplementary Material. For simplicity, we denote the probability as $P(m)$ where $m \in \{0,1,2,3...\}$, then the reduced equations of motion of the system have the simple solution as follows:
\begin{align}
P(m) =P(0)\ \left(1+\frac{1}{\bar{n}_s}\right)^{m} \dfrac{ \left(\mathcal{N}-m\right)_{(m)}}{\left(\mathcal{N}+\mathcal{D}-m\right)_{(m)}}\,,
\label{Pr}
\end{align}
where $(q)_{(m)}$ is the rising Pochhammer symbol, $\mathcal{N}$ is defined as $\mathcal{N}= \langle N\rangle+1+\dfrac{\phi \bar{n}_s+\Lambda n_{e}}{\chi (\bar{n}_s+1)}$ and $\mathcal{D}=D+\dfrac{\phi (\bar{n}_s+1)+\Lambda (n_e+1)}{\chi \bar{n}_s}-\dfrac{\phi \bar{n}_s+\Lambda n_{e}}{\chi (\bar{n}_s+1)}-1$. The probability of zero condensate phonon $P(0)$ can be determined by the normalization condition $\sum_m P(m)=1$, which gives:
\begin{align}
    P(0)^{-1}=\, _2F_1\left(1,-\mathcal{N};-\mathcal{D}-\mathcal{N}+1;1+\frac{1}{\bar{n}_s}\right)\,,
\end{align}
where $_2F_1\left(1,-\mathcal{N};-\mathcal{D}-\mathcal{N}+1;1+\dfrac{1}{\bar{n}_s}\right)$ is the Gauss hypergeometric function of type $(2,1)$. As a reminder, $D$ is the total number of oscillators in the system and $\langle N \rangle$ is the total number of phonons in all modes given in Eq.~\eqref{cn}. Examples of such distributions are shown in Supplementary Material. 

The mean phonon numbers in the condensate can be calculated by the summation $\langle n_0\rangle=\sum_m mP(m)$. The fraction of the phonon condensate $\langle \bar{n}_s \rangle/ \langle N\rangle $ is given as follows:
\begin{align}
   \frac{\langle n_0\rangle}{\langle N\rangle}=\frac{\mathcal{N} (1+\frac{1}{\bar{n}_s}) \, _2F_1(2,1-\mathcal{N};-\mathcal{D}-\mathcal{N}+2;1+\frac{1}{\bar{n}_s})}{ (\mathcal{D}+\mathcal{N}-1) \, _2F_1\left(1,-\mathcal{N};-\mathcal{D}-\mathcal{N}+1;1+\frac{1}{\bar{n}_s}\right)}/\langle N\rangle\,.
   \label{fra}
\end{align}
In deriving Eq.~\ref{Pr}-\eqref{fra}, we have treated the total phonon number as its mean value $\langle N\rangle$ given in Eq.~\eqref{cn}. The second moment of condensate phonon number takes the analytic form:
\begin{align}
    \langle n_0^2\rangle =\frac{\mathcal{N} (1+\frac{1}{\bar{n}_s}) \, _3F_2(2,1-\mathcal{N};1,-\mathcal{D}-\mathcal{N}+2;1+\frac{1}{\bar{n}_s})}{(\mathcal{D}+\mathcal{N}-1)\,_2F_1\left(1,-\mathcal{N};-\mathcal{D}-\mathcal{N}+1;1+\frac{1}{\bar{n}_s}\right)}\,.
\end{align}
The Mandel Q parameter defined by 
\begin{align}
    Q=\frac{\mathrm{Var}\,[n_0]}{\mathbb{E}\,[n_0]}-1=({\langle n_0^2\rangle -\langle n_0 \rangle^2})/{\langle n_0 \rangle} -1
\end{align} 
can be directly computed from the above expressions [see Fig.~\ref{Fig1}].

\textit{Nonequilibrium phase transition.}--
For the Gauss hypergeometric function given in Eq.~\eqref{fra}, the function $_2F_1(a,-b;-c;d)$ for positive $\{a,b,c,d\}$ is bounded when the ratio $\frac{c}{b}<d$ and it converges to a finite positive value as $b\rightarrow \infty$. Notice that in the large-$N$ limit, the total phonon number and the total vibrational modes are related by $D \simeq \dfrac{1}{(\bar{n}_s +\delta n)}\langle N\rangle$ where $\delta n=\dfrac{\Lambda}{\phi +\Lambda}(n_{ex}-\bar{n}_s)$. Therefore, in this case the fraction of condensate phonons is reduced to:
\begin{align}
   \dfrac{\langle n_0\rangle}{\langle N\rangle}&\simeq \frac{(1+\frac{1}{\bar{n}_s})}{(1+\frac{1}{\bar{n}_s+\delta n})}\frac{_2F_1\left(2,-\langle N\rangle;-(1+\frac{1}{\bar{n}_s+\delta n})\langle N\rangle;1+\frac{1}{\bar{n}_s}\right)}{_2F_1\left(1,-\langle N\rangle;-(1+\frac{1}{\bar{n}_s+\delta n})\langle N\rangle;1+\frac{1}{\bar{n}_s}\right)}/\langle N\rangle\,.
\end{align}
It can be shown that for large integer values of $\langle N\rangle$ and $\delta n <0$, the ratio of the hypergeometric functions is finite and the fraction vanishes:
\begin{align}
      &\frac{\langle n_0\rangle}{\langle N\rangle} \xrightarrow{D \rightarrow \infty} 0 \quad \mathrm{for}\ \ \delta n <0 \,\, \mathrm{and} \,\, \langle N\rangle\in \mathbb{Z^+}\,\nonumber\\
      &\frac{d}{d\,n_{ex}}\left(\frac{\langle n_0\rangle}{\langle N\rangle}\right) \xrightarrow{D \rightarrow \infty} 0 \quad \mathrm{for}\ \ \delta n <0 \,\, \mathrm{and} \,\, \langle N\rangle\in \mathbb{Z^+}\,.
\label{largen}
\end{align}
For the case $\delta n >0$, the hypergeometric functions are unbounded as $b$ goes to infinity and the ratio $_2F_1(2,-b;-c;d)/_2F_1(1,-b;-c;d)$ scales as $b$. This leads to:
\begin{align}
    &\frac{\langle n_0\rangle}{\langle N\rangle}>0 \quad \mathrm{for}\ \ \delta n>0 \,\, \mathrm{and}\ \,\, \langle N\rangle\in \mathbb{Z^+}\nonumber\\
    &\frac{d}{dn_{ex}}\left.\left(\frac{\langle n_0\rangle}{\langle N\rangle}\right)\right|_{\delta n=0^+}>0 \quad \mathrm{for}\ \ \langle N\rangle\in \mathbb{Z^+}\,.
    \label{condition1}
\end{align}
One can verify easily that the limit $\underset{\delta n\to 0}{\lim}\langle n_0 \rangle /\langle N\rangle=0$ in the large-$D$ limit and the derivative is discontinuous at the transition point $\delta n=0$. This confirms that as the external source intensifies, the NESS of the system goes through a continuous phase transition, similar to the second-order phase transition in the equilibrium superfluidity, to the condensate phase. Near the transition point, i.e. $0<(n_{ex}-\bar{n}_s)/(\bar{n}_s)\ll 1 $, we apply the leading-order saddle-point approximation and obtain the fraction of the condensate:
\begin{align}
      \frac{\langle n_0\rangle}{\langle N\rangle} \simeq \frac{1}{\bar{n}_s}\dfrac{\Lambda}{\phi +\Lambda}(n_{ex}-\bar{n}_s)\,.
 \end{align}
This confirms the result obtained by applying the decorrelation approximation Eq.~\eqref{semi_condition} and the critical exponent $\beta=1$.

%%Fig 2
 \begin{figure}[t]
    \centering
    \includegraphics[width=0.47\textwidth]{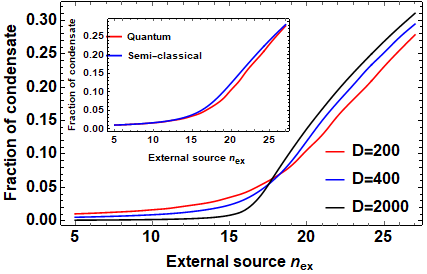}
    \caption{Fractions of condensate for different values of $D$ demonstrate the second-order phase transition at the critical external occupation number. Parameters are $\bar{n}_s=16,\, \phi=5\mathrm{GHz}$, $\Lambda=10\mathrm{GHz}$, $\chi=0.07\mathrm{GHz}$. Inset: Comparison between the decorrelation solution Eq.~\eqref{semi_solution} (semi-classical) and the solution without the deccorelation assumption (quantum) according to Eq.~\eqref{fra} for $D=200$.}
    \label{Fig2}
\end{figure}

For finite values of $D$, an equivalent condition for the condensation to occur, besides directly computing the fraction of condensation in Eq.~\eqref{fra}, is to look at the statistics of the condensate phonons, namely, when $P(n_{cr}+1)>P(n_{cr})$ for some $n_{cr}\ge 1$. This inequality gives the condensation condition
\begin{align}
    n_{ex}>\frac{(D-1)\Lambda-2\phi}{(D+1)\Lambda}\bar{n}_s+\left(\frac{\phi+\Lambda}{\chi}+n_{cr}-1\right)\frac{\phi+\Lambda}{(D+1)\Lambda}\,.
    \label{Sec2:condition}
\end{align}
The transition point can be directly computed from Eq.~\eqref{Sec2:condition} by setting $n_{cr}=1$. Notice that in the large-$D$ limit, the critical number density reduces to the same as the surrounding solvent:
\begin{align}
     n_{ex}=\bar{n}_s\,,
\end{align}
which is the same as the condition given in Eq.~\eqref{semi_condition} and Eq.~\eqref{condition1}.
In contrast with the cold atom equilibrium BEC, the Fr\"ohlich condensate is a result of the nonequilibrium driving. In this case, the pump plays a role of setting the number of particles, and the medium plays a role of setting the temperature. Therefore, BEC either arises by reducing the medium temperature at fixed pump (equilibrium case), or by increasing the pump at fixed medium temperature (nonequilibrium case). On the other hand, the emergence of the condensation also infers the off-diagonal long-range order in the system indicated by the Penrose-Onsager criterion \cite{penrose1956bose,yang1962concept}. In the long-distance limit, the single-particle density matrix becomes $\rho(x,x')=\langle \psi^\dag (x) \psi(x')\rangle \to n_c$ where $n_c$ stands for the density of the condensate. The emergence of the condensate, though obtainable from the dequantized or semi-classical equations, suggests a strong enhancement of non-classicality of the system.

\begin{figure}[t]
    \centering
    \includegraphics[width=0.45\textwidth]{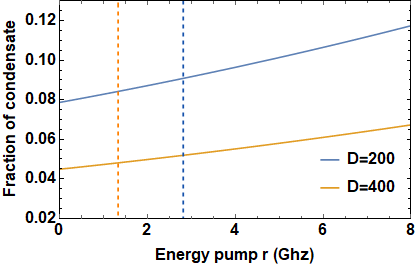}
    \caption{ Fraction of condensate vs energy input for classical pumping field. The vertical lines are the threshold pumping values. For larger $D$ values, the behaviors are similar but the threshold values are much closer to the vertical axis $r=0$. Parameters are $\bar{n}_s=16,\, \phi=5\mathrm{GHz}$, $\chi=0.07\mathrm{GHz}$.}
    \label{Fig3}
\end{figure}

The phase behavior was studied in the case of a classical pumping field  \cite{zhang2019quantum}, where the pump coupling $V_p=\sum_s \left(F_s^*(t) a_s e^{-i\omega_s t}+\mathrm{h.c.}\right)$ is featured by a broad spectrum $\langle F_s(t')F_s^*(t)\rangle =\frac{r}{2}\delta(t-t')$ with the control parameter $r$ denoting the pumping rate. However, the issue with the classical pumping field is that it is qualitatively similar to an external source at an infinitely high temperature in that the emission and absorption rates are the same in a microprocess. The only parameter describing such classical pumps is the pumping rate $r$. Since the key to determine the phase transition is the comparison of the pump particle density with that of the bath, this requires a quantum description of not only the bath but also the pump. By treating the pump classically, all modes obtain the same gaining rate due to the pump regardless of the number of phonons already on the modes. Therein many of the critical properties at low temperatures and near the transition are erased. For instance, the critical energy supply at which the condensation emerges, in this situation, approaches zero in the large-$D$ limit: $r=\frac{\phi}{D+1}(1+\frac{\phi}{\chi})\xrightarrow{D\rightarrow \infty}0$. It means that the condensate phonons at the lowest mode start to accumulate from the zero pumping and increase with the strength of the pumping field until saturation. It can be further shown that the fraction of the condensation in the large-$D$ limit is roughly $\dfrac{r/\phi}{ r/\phi+\bar n_s}$. As shown in Fig.~\ref{Fig2} and Fig.~\ref{Fig3}, the condensation will emerge regardless the types of irradiating energy, but the transition to the condensation phase is a continuous crossover with no real distinction between the ``phases" in this case [Fig.~\ref{Fig3}].

In Fig.~\ref{Fig1}, we show that the Mandel $Q$ parameter reaches the maximal at the transition point near $n_{ex}=\bar{n}_s=16$ (the exact value is given in Eq.~\eqref{Sec2:condition}) and decays rapidly away from the phase transition point. For extremely large energy input, $Q$ becomes negative and it suggests a sub-Poissonian statistics similar to that in the resonance fluorescence of a single atom. In Fig.~\ref{Fig2}, the fraction of condensate phonons as the occupation of the external source is simulated. For relatively small total phonon numbers (e.g. $\langle N\rangle=200$), the transition from the ordinary nonequilibrium phase to the condensation phase is similar to a continuous crossover due to the finiteness of the system. In particular, in this case the condensation fraction is nonzero even for zero external energy input due to the finiteness of the system and the thermal equilibrium with the surounding solvent. However, in the large-$D$ case, the transition becomes sharp and the fraction is negligible at zero input, which is expected for the thermodynamic limit. This agrees with the findings reported in a recent experiment \cite{nardecchia2018out}. We compare the solution with the decorrelation approximation Eq.~\eqref{semi_solution} and the solution without the assumption Eq.~\eqref{fra}. The relative difference maximizes near the transition point (for $D=200$, the critical point is at $n_{ex}\simeq 17.4$). The numerical simulations verifies that the transition behaves in accordance with our analytical result Eq.~\eqref{Sec2:condition}. In Fig.~\ref{Fig3}, we show that in the case of a classical broad-spectrum pumping, such transition does not show up.

\textit{Conclusions.}--It has been long debated if molecules go through certain phase transitions to the condensate states during the Fr\"ohlich process. In this Letter, we derived a full quantum theory of the Fr\"ohlich condensation and demonstrated the critical phenomenon. In particular, we presented an analytical proof of the existence of the nonequilibrium phase transition based on Wu-Austin Hamiltonian and outlined the conditions for witnessing such transitions. We analyzed the phonon distributions and fluctuations near the transition and showed that the phase transition is only realizable in the large-$D$ limit. In biological systems, this condition can be naturally realized and the analysis demonstrates the viability of broad experimental prospect to observe such nonequilibrium phase transition of the molecular vibrations into a coherent quantum state in much broader systems than the equilibrium BEC. Recent experiments have made tremendous progress in the nonequilibrium condensation in biological systems, and signs of such phase transition as shown in Fig.~\ref{Fig2} was witnessed for the first time. Further experimental investigations and smoking-gun evidences are in need to exclude all potential artifacts and to confirm the finding. It is hopeful that this will open a new door for further explorations of out-of equilibrium collective oscillations in broader nonequilibrium systems.

X.W wants to thank Zhedong Zhang for helpful discussions and acknowledge the support from grant WIUCASQD2022026.
\bibliography{bib}

\providecommand{\noopsort}[1]{}\providecommand{\singleletter}[1]{#1}%
\begin{thebibliography}{46}%
\makeatletter
\providecommand \@ifxundefined [1]{%
 \@ifx{#1\undefined}
}%
\providecommand \@ifnum [1]{%
 \ifnum #1\expandafter \@firstoftwo
 \else \expandafter \@secondoftwo
 \fi
}%
\providecommand \@ifx [1]{%
 \ifx #1\expandafter \@firstoftwo
 \else \expandafter \@secondoftwo
 \fi
}%
\providecommand \natexlab [1]{#1}%
\providecommand \enquote  [1]{``#1''}%
\providecommand \bibnamefont  [1]{#1}%
\providecommand \bibfnamefont [1]{#1}%
\providecommand \citenamefont [1]{#1}%
\providecommand \href@noop [0]{\@secondoftwo}%
\providecommand \href [0]{\begingroup \@sanitize@url \@href}%
\providecommand \@href[1]{\@@startlink{#1}\@@href}%
\providecommand \@@href[1]{\endgroup#1\@@endlink}%
\providecommand \@sanitize@url [0]{\catcode `\\12\catcode `\$12\catcode
  `\&12\catcode `\#12\catcode `\^12\catcode `\_12\catcode `\%12\relax}%
\providecommand \@@startlink[1]{}%
\providecommand \@@endlink[0]{}%
\providecommand \url  [0]{\begingroup\@sanitize@url \@url }%
\providecommand \@url [1]{\endgroup\@href {#1}{\urlprefix }}%
\providecommand \urlprefix  [0]{URL }%
\providecommand \Eprint [0]{\href }%
\providecommand \doibase [0]{https://doi.org/}%
\providecommand \selectlanguage [0]{\@gobble}%
\providecommand \bibinfo  [0]{\@secondoftwo}%
\providecommand \bibfield  [0]{\@secondoftwo}%
\providecommand \translation [1]{[#1]}%
\providecommand \BibitemOpen [0]{}%
\providecommand \bibitemStop [0]{}%
\providecommand \bibitemNoStop [0]{.\EOS\space}%
\providecommand \EOS [0]{\spacefactor3000\relax}%
\providecommand \BibitemShut  [1]{\csname bibitem#1\endcsname}%
\let\auto@bib@innerbib\@empty
%</preamble>
\bibitem [{\citenamefont
  {Fr{\"o}hlich}(1968{\natexlab{a}})}]{frohlich1968long}%
  \BibitemOpen
  \bibfield  {author} {\bibinfo {author} {\bibfnamefont {H.}~\bibnamefont
  {Fr{\"o}hlich}},\ }\bibfield  {title} {\bibinfo {title} {Long-range coherence
  and energy storage in biological systems},\ }\href@noop {} {\bibfield
  {journal} {\bibinfo  {journal} {International Journal of Quantum Chemistry}\
  }\textbf {\bibinfo {volume} {2}},\ \bibinfo {pages} {641} (\bibinfo {year}
  {1968}{\natexlab{a}})}\BibitemShut {NoStop}%
\bibitem [{\citenamefont
  {Fr{\"o}hlich}(1968{\natexlab{b}})}]{frohlich1968bose}%
  \BibitemOpen
  \bibfield  {author} {\bibinfo {author} {\bibfnamefont {H.}~\bibnamefont
  {Fr{\"o}hlich}},\ }\bibfield  {title} {\bibinfo {title} {Bose condensation of
  strongly excited longitudinal electric modes},\ }\href@noop {} {\bibfield
  {journal} {\bibinfo  {journal} {Physics Letters A}\ }\textbf {\bibinfo
  {volume} {26}},\ \bibinfo {pages} {402} (\bibinfo {year}
  {1968}{\natexlab{b}})}\BibitemShut {NoStop}%
\bibitem [{\citenamefont {Fr{\"o}hlich}(1970)}]{frohlich1970long}%
  \BibitemOpen
  \bibfield  {author} {\bibinfo {author} {\bibfnamefont {H.}~\bibnamefont
  {Fr{\"o}hlich}},\ }\bibfield  {title} {\bibinfo {title} {Long range coherence
  and the action of enzymes},\ }\href@noop {} {\bibfield  {journal} {\bibinfo
  {journal} {Nature}\ }\textbf {\bibinfo {volume} {228}},\ \bibinfo {pages}
  {1093} (\bibinfo {year} {1970})}\BibitemShut {NoStop}%
\bibitem [{\citenamefont {Fr{\"o}hlich}(1988)}]{Frohlich1988}%
  \BibitemOpen
  \bibfield  {author} {\bibinfo {author} {\bibfnamefont {H.}~\bibnamefont
  {Fr{\"o}hlich}},\ }\bibinfo {title} {Theoretical physics and biology},\ in\
  \href {https://doi.org/10.1007/978-3-642-73309-3_1} {\emph {\bibinfo
  {booktitle} {Biological Coherence and Response to External Stimuli}}},\
  \bibinfo {editor} {edited by\ \bibinfo {editor} {\bibfnamefont
  {H.}~\bibnamefont {Fr{\"o}hlich}}}\ (\bibinfo  {publisher} {Springer Berlin
  Heidelberg},\ \bibinfo {address} {Berlin, Heidelberg},\ \bibinfo {year}
  {1988})\ pp.\ \bibinfo {pages} {1--24}\BibitemShut {NoStop}%
\bibitem [{\citenamefont {Anderson}\ \emph {et~al.}(1995)\citenamefont
  {Anderson}, \citenamefont {Ensher}, \citenamefont {Matthews}, \citenamefont
  {Wieman},\ and\ \citenamefont {Cornell}}]{anderson1995observation}%
  \BibitemOpen
  \bibfield  {author} {\bibinfo {author} {\bibfnamefont {M.~H.}\ \bibnamefont
  {Anderson}}, \bibinfo {author} {\bibfnamefont {J.~R.}\ \bibnamefont
  {Ensher}}, \bibinfo {author} {\bibfnamefont {M.~R.}\ \bibnamefont
  {Matthews}}, \bibinfo {author} {\bibfnamefont {C.~E.}\ \bibnamefont
  {Wieman}},\ and\ \bibinfo {author} {\bibfnamefont {E.~A.}\ \bibnamefont
  {Cornell}},\ }\bibfield  {title} {\bibinfo {title} {Observation of
  bose-einstein condensation in a dilute atomic vapor},\ }\href@noop {}
  {\bibfield  {journal} {\bibinfo  {journal} {science}\ }\textbf {\bibinfo
  {volume} {269}},\ \bibinfo {pages} {198} (\bibinfo {year}
  {1995})}\BibitemShut {NoStop}%
\bibitem [{\citenamefont {Davis}\ \emph {et~al.}(1995)\citenamefont {Davis},
  \citenamefont {Mewes}, \citenamefont {Andrews}, \citenamefont {van Druten},
  \citenamefont {Durfee}, \citenamefont {Kurn},\ and\ \citenamefont
  {Ketterle}}]{davis1995bose}%
  \BibitemOpen
  \bibfield  {author} {\bibinfo {author} {\bibfnamefont {K.~B.}\ \bibnamefont
  {Davis}}, \bibinfo {author} {\bibfnamefont {M.-O.}\ \bibnamefont {Mewes}},
  \bibinfo {author} {\bibfnamefont {M.~R.}\ \bibnamefont {Andrews}}, \bibinfo
  {author} {\bibfnamefont {N.~J.}\ \bibnamefont {van Druten}}, \bibinfo
  {author} {\bibfnamefont {D.~S.}\ \bibnamefont {Durfee}}, \bibinfo {author}
  {\bibfnamefont {D.}~\bibnamefont {Kurn}},\ and\ \bibinfo {author}
  {\bibfnamefont {W.}~\bibnamefont {Ketterle}},\ }\bibfield  {title} {\bibinfo
  {title} {Bose-einstein condensation in a gas of sodium atoms},\ }\href@noop
  {} {\bibfield  {journal} {\bibinfo  {journal} {Physical review letters}\
  }\textbf {\bibinfo {volume} {75}},\ \bibinfo {pages} {3969} (\bibinfo {year}
  {1995})}\BibitemShut {NoStop}%
\bibitem [{\citenamefont {Bunkov}\ and\ \citenamefont
  {Volovik}(2013)}]{bunkov2013spin}%
  \BibitemOpen
  \bibfield  {author} {\bibinfo {author} {\bibfnamefont {Y.~M.}\ \bibnamefont
  {Bunkov}}\ and\ \bibinfo {author} {\bibfnamefont {G.}~\bibnamefont
  {Volovik}},\ }\bibfield  {title} {\bibinfo {title} {Spin superfluidity and
  magnon bose-einstein condensation},\ }\href@noop {} {\bibfield  {journal}
  {\bibinfo  {journal} {KH Bennemann and JB Ketterson, International Series of
  Monographs on Physics}\ }\textbf {\bibinfo {volume} {156}},\ \bibinfo {pages}
  {253} (\bibinfo {year} {2013})}\BibitemShut {NoStop}%
\bibitem [{\citenamefont {Demokritov}\ \emph {et~al.}(2006)\citenamefont
  {Demokritov}, \citenamefont {Demidov}, \citenamefont {Dzyapko}, \citenamefont
  {Melkov}, \citenamefont {Serga}, \citenamefont {Hillebrands},\ and\
  \citenamefont {Slavin}}]{demokritov2006bose}%
  \BibitemOpen
  \bibfield  {author} {\bibinfo {author} {\bibfnamefont {S.~O.}\ \bibnamefont
  {Demokritov}}, \bibinfo {author} {\bibfnamefont {V.~E.}\ \bibnamefont
  {Demidov}}, \bibinfo {author} {\bibfnamefont {O.}~\bibnamefont {Dzyapko}},
  \bibinfo {author} {\bibfnamefont {G.~A.}\ \bibnamefont {Melkov}}, \bibinfo
  {author} {\bibfnamefont {A.~A.}\ \bibnamefont {Serga}}, \bibinfo {author}
  {\bibfnamefont {B.}~\bibnamefont {Hillebrands}},\ and\ \bibinfo {author}
  {\bibfnamefont {A.~N.}\ \bibnamefont {Slavin}},\ }\bibfield  {title}
  {\bibinfo {title} {Bose--einstein condensation of quasi-equilibrium magnons
  at room temperature under pumping},\ }\href@noop {} {\bibfield  {journal}
  {\bibinfo  {journal} {Nature}\ }\textbf {\bibinfo {volume} {443}},\ \bibinfo
  {pages} {430} (\bibinfo {year} {2006})}\BibitemShut {NoStop}%
\bibitem [{\citenamefont {Chumak}\ \emph {et~al.}(2009)\citenamefont {Chumak},
  \citenamefont {Melkov}, \citenamefont {Demidov}, \citenamefont {Dzyapko},
  \citenamefont {Safonov},\ and\ \citenamefont {Demokritov}}]{chumak2009bose}%
  \BibitemOpen
  \bibfield  {author} {\bibinfo {author} {\bibfnamefont {A.}~\bibnamefont
  {Chumak}}, \bibinfo {author} {\bibfnamefont {G.}~\bibnamefont {Melkov}},
  \bibinfo {author} {\bibfnamefont {V.}~\bibnamefont {Demidov}}, \bibinfo
  {author} {\bibfnamefont {O.}~\bibnamefont {Dzyapko}}, \bibinfo {author}
  {\bibfnamefont {V.}~\bibnamefont {Safonov}},\ and\ \bibinfo {author}
  {\bibfnamefont {S.}~\bibnamefont {Demokritov}},\ }\bibfield  {title}
  {\bibinfo {title} {Bose-einstein condensation of magnons under incoherent
  pumping},\ }\href@noop {} {\bibfield  {journal} {\bibinfo  {journal}
  {Physical review letters}\ }\textbf {\bibinfo {volume} {102}},\ \bibinfo
  {pages} {187205} (\bibinfo {year} {2009})}\BibitemShut {NoStop}%
\bibitem [{\citenamefont {Kasprzak}\ \emph {et~al.}(2006)\citenamefont
  {Kasprzak}, \citenamefont {Richard}, \citenamefont {Kundermann},
  \citenamefont {Baas}, \citenamefont {Jeambrun}, \citenamefont {Keeling},
  \citenamefont {Marchetti}, \citenamefont {Szyma{\'n}ska}, \citenamefont
  {Andr{\'e}}, \citenamefont {Staehli} \emph {et~al.}}]{kasprzak2006bose}%
  \BibitemOpen
  \bibfield  {author} {\bibinfo {author} {\bibfnamefont {J.}~\bibnamefont
  {Kasprzak}}, \bibinfo {author} {\bibfnamefont {M.}~\bibnamefont {Richard}},
  \bibinfo {author} {\bibfnamefont {S.}~\bibnamefont {Kundermann}}, \bibinfo
  {author} {\bibfnamefont {A.}~\bibnamefont {Baas}}, \bibinfo {author}
  {\bibfnamefont {P.}~\bibnamefont {Jeambrun}}, \bibinfo {author}
  {\bibfnamefont {J.~M.~J.}\ \bibnamefont {Keeling}}, \bibinfo {author}
  {\bibfnamefont {F.}~\bibnamefont {Marchetti}}, \bibinfo {author}
  {\bibfnamefont {M.}~\bibnamefont {Szyma{\'n}ska}}, \bibinfo {author}
  {\bibfnamefont {R.}~\bibnamefont {Andr{\'e}}}, \bibinfo {author}
  {\bibfnamefont {J.}~\bibnamefont {Staehli}}, \emph {et~al.},\ }\bibfield
  {title} {\bibinfo {title} {Bose--einstein condensation of exciton
  polaritons},\ }\href@noop {} {\bibfield  {journal} {\bibinfo  {journal}
  {Nature}\ }\textbf {\bibinfo {volume} {443}},\ \bibinfo {pages} {409}
  (\bibinfo {year} {2006})}\BibitemShut {NoStop}%
\bibitem [{\citenamefont {Manni}\ \emph {et~al.}(2011)\citenamefont {Manni},
  \citenamefont {Lagoudakis}, \citenamefont {Pietka}, \citenamefont
  {Fontanesi}, \citenamefont {Wouters}, \citenamefont {Savona}, \citenamefont
  {Andr{\'e}},\ and\ \citenamefont {Deveaud-Pl{\'e}dran}}]{manni2011polariton}%
  \BibitemOpen
  \bibfield  {author} {\bibinfo {author} {\bibfnamefont {F.}~\bibnamefont
  {Manni}}, \bibinfo {author} {\bibfnamefont {K.~G.}\ \bibnamefont
  {Lagoudakis}}, \bibinfo {author} {\bibfnamefont {B.}~\bibnamefont {Pietka}},
  \bibinfo {author} {\bibfnamefont {L.}~\bibnamefont {Fontanesi}}, \bibinfo
  {author} {\bibfnamefont {M.}~\bibnamefont {Wouters}}, \bibinfo {author}
  {\bibfnamefont {V.}~\bibnamefont {Savona}}, \bibinfo {author} {\bibfnamefont
  {R.}~\bibnamefont {Andr{\'e}}},\ and\ \bibinfo {author} {\bibfnamefont
  {B.}~\bibnamefont {Deveaud-Pl{\'e}dran}},\ }\bibfield  {title} {\bibinfo
  {title} {Polariton condensation in a one-dimensional disordered potential},\
  }\href@noop {} {\bibfield  {journal} {\bibinfo  {journal} {Physical Review
  Letters}\ }\textbf {\bibinfo {volume} {106}},\ \bibinfo {pages} {176401}
  (\bibinfo {year} {2011})}\BibitemShut {NoStop}%
\bibitem [{\citenamefont {Deng}\ \emph {et~al.}(2010)\citenamefont {Deng},
  \citenamefont {Haug},\ and\ \citenamefont {Yamamoto}}]{deng2010exciton}%
  \BibitemOpen
  \bibfield  {author} {\bibinfo {author} {\bibfnamefont {H.}~\bibnamefont
  {Deng}}, \bibinfo {author} {\bibfnamefont {H.}~\bibnamefont {Haug}},\ and\
  \bibinfo {author} {\bibfnamefont {Y.}~\bibnamefont {Yamamoto}},\ }\bibfield
  {title} {\bibinfo {title} {Exciton-polariton bose-einstein condensation},\
  }\href@noop {} {\bibfield  {journal} {\bibinfo  {journal} {Reviews of modern
  physics}\ }\textbf {\bibinfo {volume} {82}},\ \bibinfo {pages} {1489}
  (\bibinfo {year} {2010})}\BibitemShut {NoStop}%
\bibitem [{\citenamefont {Zhang}\ \emph {et~al.}(2022)\citenamefont {Zhang},
  \citenamefont {Zhao},\ and\ \citenamefont {Lei}}]{zhang2022quantum}%
  \BibitemOpen
  \bibfield  {author} {\bibinfo {author} {\bibfnamefont {Z.}~\bibnamefont
  {Zhang}}, \bibinfo {author} {\bibfnamefont {S.}~\bibnamefont {Zhao}},\ and\
  \bibinfo {author} {\bibfnamefont {D.}~\bibnamefont {Lei}},\ }\bibfield
  {title} {\bibinfo {title} {Quantum fluctuations and coherence of a molecular
  polariton condensate},\ }\href@noop {} {\bibfield  {journal} {\bibinfo
  {journal} {arXiv preprint arXiv:2204.13528}\ } (\bibinfo {year}
  {2022})}\BibitemShut {NoStop}%
\bibitem [{\citenamefont {Bloch}\ \emph {et~al.}(2022)\citenamefont {Bloch},
  \citenamefont {Carusotto},\ and\ \citenamefont {Wouters}}]{bloch2022non}%
  \BibitemOpen
  \bibfield  {author} {\bibinfo {author} {\bibfnamefont {J.}~\bibnamefont
  {Bloch}}, \bibinfo {author} {\bibfnamefont {I.}~\bibnamefont {Carusotto}},\
  and\ \bibinfo {author} {\bibfnamefont {M.}~\bibnamefont {Wouters}},\
  }\bibfield  {title} {\bibinfo {title} {Non-equilibrium bose--einstein
  condensation in photonic systems},\ }\href@noop {} {\bibfield  {journal}
  {\bibinfo  {journal} {Nature Reviews Physics}\ ,\ \bibinfo {pages} {1}}
  (\bibinfo {year} {2022})}\BibitemShut {NoStop}%
\bibitem [{\citenamefont {Haken}(1975)}]{haken1975cooperative}%
  \BibitemOpen
  \bibfield  {author} {\bibinfo {author} {\bibfnamefont {H.}~\bibnamefont
  {Haken}},\ }\bibfield  {title} {\bibinfo {title} {Cooperative phenomena in
  systems far from thermal equilibrium and in nonphysical systems},\
  }\href@noop {} {\bibfield  {journal} {\bibinfo  {journal} {Reviews of modern
  physics}\ }\textbf {\bibinfo {volume} {47}},\ \bibinfo {pages} {67} (\bibinfo
  {year} {1975})}\BibitemShut {NoStop}%
\bibitem [{\citenamefont {Haken}(1977)}]{haken1977synergetics}%
  \BibitemOpen
  \bibfield  {author} {\bibinfo {author} {\bibfnamefont {H.}~\bibnamefont
  {Haken}},\ }\href@noop {} {\emph {\bibinfo {title} {Synergetics. An
  Introduction. Nonequilibrium Phase Trasitions and Self-organization in
  Physics, Chemistry, and Biology}}}\ (\bibinfo  {publisher} {Springer},\
  \bibinfo {year} {1977})\BibitemShut {NoStop}%
\bibitem [{\citenamefont {Pokorn{\`y}}(2004)}]{pokorny2004excitation}%
  \BibitemOpen
  \bibfield  {author} {\bibinfo {author} {\bibfnamefont {J.}~\bibnamefont
  {Pokorn{\`y}}},\ }\bibfield  {title} {\bibinfo {title} {Excitation of
  vibrations in microtubules in living cells},\ }\href@noop {} {\bibfield
  {journal} {\bibinfo  {journal} {Bioelectrochemistry}\ }\textbf {\bibinfo
  {volume} {63}},\ \bibinfo {pages} {321} (\bibinfo {year} {2004})}\BibitemShut
  {NoStop}%
\bibitem [{\citenamefont {Reimers}\ \emph {et~al.}(2009)\citenamefont
  {Reimers}, \citenamefont {McKemmish}, \citenamefont {McKenzie}, \citenamefont
  {Mark},\ and\ \citenamefont {Hush}}]{reimers2009weak}%
  \BibitemOpen
  \bibfield  {author} {\bibinfo {author} {\bibfnamefont {J.~R.}\ \bibnamefont
  {Reimers}}, \bibinfo {author} {\bibfnamefont {L.~K.}\ \bibnamefont
  {McKemmish}}, \bibinfo {author} {\bibfnamefont {R.~H.}\ \bibnamefont
  {McKenzie}}, \bibinfo {author} {\bibfnamefont {A.~E.}\ \bibnamefont {Mark}},\
  and\ \bibinfo {author} {\bibfnamefont {N.~S.}\ \bibnamefont {Hush}},\
  }\bibfield  {title} {\bibinfo {title} {Weak, strong, and coherent regimes of
  fr{\"o}hlich condensation and their applications to terahertz medicine and
  quantum consciousness},\ }\href@noop {} {\bibfield  {journal} {\bibinfo
  {journal} {Proceedings of the National Academy of Sciences}\ }\textbf
  {\bibinfo {volume} {106}},\ \bibinfo {pages} {4219} (\bibinfo {year}
  {2009})}\BibitemShut {NoStop}%
\bibitem [{\citenamefont {Turton}\ \emph {et~al.}(2014)\citenamefont {Turton},
  \citenamefont {Senn}, \citenamefont {Harwood}, \citenamefont {Lapthorn},
  \citenamefont {Ellis},\ and\ \citenamefont {Wynne}}]{turton2014terahertz}%
  \BibitemOpen
  \bibfield  {author} {\bibinfo {author} {\bibfnamefont {D.~A.}\ \bibnamefont
  {Turton}}, \bibinfo {author} {\bibfnamefont {H.~M.}\ \bibnamefont {Senn}},
  \bibinfo {author} {\bibfnamefont {T.}~\bibnamefont {Harwood}}, \bibinfo
  {author} {\bibfnamefont {A.~J.}\ \bibnamefont {Lapthorn}}, \bibinfo {author}
  {\bibfnamefont {E.~M.}\ \bibnamefont {Ellis}},\ and\ \bibinfo {author}
  {\bibfnamefont {K.}~\bibnamefont {Wynne}},\ }\bibfield  {title} {\bibinfo
  {title} {Terahertz underdamped vibrational motion governs protein-ligand
  binding in solution},\ }\href@noop {} {\bibfield  {journal} {\bibinfo
  {journal} {Nature communications}\ }\textbf {\bibinfo {volume} {5}},\
  \bibinfo {pages} {1} (\bibinfo {year} {2014})}\BibitemShut {NoStop}%
\bibitem [{\citenamefont {Lundholm}\ \emph {et~al.}(2015)\citenamefont
  {Lundholm}, \citenamefont {Rodilla}, \citenamefont {Wahlgren}, \citenamefont
  {Duelli}, \citenamefont {Bourenkov}, \citenamefont {Vukusic}, \citenamefont
  {Friedman}, \citenamefont {Stake}, \citenamefont {Schneider},\ and\
  \citenamefont {Katona}}]{lundholm2015terahertz}%
  \BibitemOpen
  \bibfield  {author} {\bibinfo {author} {\bibfnamefont {I.~V.}\ \bibnamefont
  {Lundholm}}, \bibinfo {author} {\bibfnamefont {H.}~\bibnamefont {Rodilla}},
  \bibinfo {author} {\bibfnamefont {W.~Y.}\ \bibnamefont {Wahlgren}}, \bibinfo
  {author} {\bibfnamefont {A.}~\bibnamefont {Duelli}}, \bibinfo {author}
  {\bibfnamefont {G.}~\bibnamefont {Bourenkov}}, \bibinfo {author}
  {\bibfnamefont {J.}~\bibnamefont {Vukusic}}, \bibinfo {author} {\bibfnamefont
  {R.}~\bibnamefont {Friedman}}, \bibinfo {author} {\bibfnamefont
  {J.}~\bibnamefont {Stake}}, \bibinfo {author} {\bibfnamefont
  {T.}~\bibnamefont {Schneider}},\ and\ \bibinfo {author} {\bibfnamefont
  {G.}~\bibnamefont {Katona}},\ }\bibfield  {title} {\bibinfo {title}
  {Terahertz radiation induces non-thermal structural changes associated with
  fr{\"o}hlich condensation in a protein crystal},\ }\href@noop {} {\bibfield
  {journal} {\bibinfo  {journal} {Structural Dynamics}\ }\textbf {\bibinfo
  {volume} {2}},\ \bibinfo {pages} {054702} (\bibinfo {year}
  {2015})}\BibitemShut {NoStop}%
\bibitem [{\citenamefont {Nardecchia}\ \emph {et~al.}(2018)\citenamefont
  {Nardecchia}, \citenamefont {Torres}, \citenamefont {Lechelon}, \citenamefont
  {Giliberti}, \citenamefont {Ortolani}, \citenamefont {Nouvel}, \citenamefont
  {Gori}, \citenamefont {Meriguet}, \citenamefont {Donato}, \citenamefont
  {Preto} \emph {et~al.}}]{nardecchia2018out}%
  \BibitemOpen
  \bibfield  {author} {\bibinfo {author} {\bibfnamefont {I.}~\bibnamefont
  {Nardecchia}}, \bibinfo {author} {\bibfnamefont {J.}~\bibnamefont {Torres}},
  \bibinfo {author} {\bibfnamefont {M.}~\bibnamefont {Lechelon}}, \bibinfo
  {author} {\bibfnamefont {V.}~\bibnamefont {Giliberti}}, \bibinfo {author}
  {\bibfnamefont {M.}~\bibnamefont {Ortolani}}, \bibinfo {author}
  {\bibfnamefont {P.}~\bibnamefont {Nouvel}}, \bibinfo {author} {\bibfnamefont
  {M.}~\bibnamefont {Gori}}, \bibinfo {author} {\bibfnamefont {Y.}~\bibnamefont
  {Meriguet}}, \bibinfo {author} {\bibfnamefont {I.}~\bibnamefont {Donato}},
  \bibinfo {author} {\bibfnamefont {J.}~\bibnamefont {Preto}}, \emph {et~al.},\
  }\bibfield  {title} {\bibinfo {title} {Out-of-equilibrium collective
  oscillation as phonon condensation in a model protein},\ }\href@noop {}
  {\bibfield  {journal} {\bibinfo  {journal} {Physical Review X}\ }\textbf
  {\bibinfo {volume} {8}},\ \bibinfo {pages} {031061} (\bibinfo {year}
  {2018})}\BibitemShut {NoStop}%
\bibitem [{\citenamefont {Zhang}\ \emph {et~al.}(2019)\citenamefont {Zhang},
  \citenamefont {Agarwal},\ and\ \citenamefont {Scully}}]{zhang2019quantum}%
  \BibitemOpen
  \bibfield  {author} {\bibinfo {author} {\bibfnamefont {Z.}~\bibnamefont
  {Zhang}}, \bibinfo {author} {\bibfnamefont {G.~S.}\ \bibnamefont {Agarwal}},\
  and\ \bibinfo {author} {\bibfnamefont {M.~O.}\ \bibnamefont {Scully}},\
  }\bibfield  {title} {\bibinfo {title} {Quantum fluctuations in the
  fr{\"o}hlich condensate of molecular vibrations driven far from
  equilibrium},\ }\href@noop {} {\bibfield  {journal} {\bibinfo  {journal}
  {Physical Review Letters}\ }\textbf {\bibinfo {volume} {122}},\ \bibinfo
  {pages} {158101} (\bibinfo {year} {2019})}\BibitemShut {NoStop}%
\bibitem [{\citenamefont {Zheng}\ and\ \citenamefont
  {Li}(2021)}]{zheng2021frohlich}%
  \BibitemOpen
  \bibfield  {author} {\bibinfo {author} {\bibfnamefont {X.}~\bibnamefont
  {Zheng}}\ and\ \bibinfo {author} {\bibfnamefont {B.}~\bibnamefont {Li}},\
  }\bibfield  {title} {\bibinfo {title} {Fr{\"o}hlich condensate of phonons in
  optomechanical systems},\ }\href@noop {} {\bibfield  {journal} {\bibinfo
  {journal} {Physical Review A}\ }\textbf {\bibinfo {volume} {104}},\ \bibinfo
  {pages} {043512} (\bibinfo {year} {2021})}\BibitemShut {NoStop}%
\bibitem [{\citenamefont {Wu}\ and\ \citenamefont {Austin}(1977)}]{wu1977bose}%
  \BibitemOpen
  \bibfield  {author} {\bibinfo {author} {\bibfnamefont {T.}~\bibnamefont
  {Wu}}\ and\ \bibinfo {author} {\bibfnamefont {S.}~\bibnamefont {Austin}},\
  }\bibfield  {title} {\bibinfo {title} {Bose condensation in biosystems},\
  }\href@noop {} {\bibfield  {journal} {\bibinfo  {journal} {Physics Letters
  A}\ }\textbf {\bibinfo {volume} {64}},\ \bibinfo {pages} {151} (\bibinfo
  {year} {1977})}\BibitemShut {NoStop}%
\bibitem [{\citenamefont {Wu}\ and\ \citenamefont
  {Austin}(1978{\natexlab{a}})}]{wu1978bose}%
  \BibitemOpen
  \bibfield  {author} {\bibinfo {author} {\bibfnamefont {T.}~\bibnamefont
  {Wu}}\ and\ \bibinfo {author} {\bibfnamefont {S.}~\bibnamefont {Austin}},\
  }\bibfield  {title} {\bibinfo {title} {Bose-einstein condensation in
  biological systems},\ }\href@noop {} {\bibfield  {journal} {\bibinfo
  {journal} {Journal of theoretical biology}\ }\textbf {\bibinfo {volume}
  {71}},\ \bibinfo {pages} {209} (\bibinfo {year}
  {1978}{\natexlab{a}})}\BibitemShut {NoStop}%
\bibitem [{\citenamefont {Wu}\ and\ \citenamefont
  {Austin}(1978{\natexlab{b}})}]{wu1978cooperative}%
  \BibitemOpen
  \bibfield  {author} {\bibinfo {author} {\bibfnamefont {T.}~\bibnamefont
  {Wu}}\ and\ \bibinfo {author} {\bibfnamefont {S.}~\bibnamefont {Austin}},\
  }\bibfield  {title} {\bibinfo {title} {Cooperative behavior in biological
  systems},\ }\href@noop {} {\bibfield  {journal} {\bibinfo  {journal} {Physics
  Letters A}\ }\textbf {\bibinfo {volume} {65}},\ \bibinfo {pages} {74}
  (\bibinfo {year} {1978}{\natexlab{b}})}\BibitemShut {NoStop}%
\bibitem [{\citenamefont {Wu}\ and\ \citenamefont
  {Austin}(1981)}]{wu1981frohlich}%
  \BibitemOpen
  \bibfield  {author} {\bibinfo {author} {\bibfnamefont {T.}~\bibnamefont
  {Wu}}\ and\ \bibinfo {author} {\bibfnamefont {S.~J.}\ \bibnamefont
  {Austin}},\ }\bibfield  {title} {\bibinfo {title} {Fr{\"o}hlich's model of
  bose condensation in biological systems},\ }\href@noop {} {\bibfield
  {journal} {\bibinfo  {journal} {Journal of Biological Physics}\ }\textbf
  {\bibinfo {volume} {9}},\ \bibinfo {pages} {97} (\bibinfo {year}
  {1981})}\BibitemShut {NoStop}%
\bibitem [{\citenamefont {Preto}(2017)}]{preto2017semi}%
  \BibitemOpen
  \bibfield  {author} {\bibinfo {author} {\bibfnamefont {J.}~\bibnamefont
  {Preto}},\ }\bibfield  {title} {\bibinfo {title} {Semi-classical statistical
  description of fr{\"o}hlich condensation},\ }\href@noop {} {\bibfield
  {journal} {\bibinfo  {journal} {Journal of biological physics}\ }\textbf
  {\bibinfo {volume} {43}},\ \bibinfo {pages} {167} (\bibinfo {year}
  {2017})}\BibitemShut {NoStop}%
\bibitem [{\citenamefont {Rice}\ and\ \citenamefont
  {Carmichael}(1994)}]{rice1994photon}%
  \BibitemOpen
  \bibfield  {author} {\bibinfo {author} {\bibfnamefont {P.~R.}\ \bibnamefont
  {Rice}}\ and\ \bibinfo {author} {\bibfnamefont {H.}~\bibnamefont
  {Carmichael}},\ }\bibfield  {title} {\bibinfo {title} {Photon statistics of a
  cavity-qed laser: A comment on the laser--phase-transition analogy},\
  }\href@noop {} {\bibfield  {journal} {\bibinfo  {journal} {Physical Review
  A}\ }\textbf {\bibinfo {volume} {50}},\ \bibinfo {pages} {4318} (\bibinfo
  {year} {1994})}\BibitemShut {NoStop}%
\bibitem [{\citenamefont {Scully}\ and\ \citenamefont
  {Lamb~Jr}(1966)}]{scully1966quantum}%
  \BibitemOpen
  \bibfield  {author} {\bibinfo {author} {\bibfnamefont {M.}~\bibnamefont
  {Scully}}\ and\ \bibinfo {author} {\bibfnamefont {W.}~\bibnamefont
  {Lamb~Jr}},\ }\bibfield  {title} {\bibinfo {title} {Quantum theory of an
  optical maser},\ }\href@noop {} {\bibfield  {journal} {\bibinfo  {journal}
  {Physical Review Letters}\ }\textbf {\bibinfo {volume} {16}},\ \bibinfo
  {pages} {853} (\bibinfo {year} {1966})}\BibitemShut {NoStop}%
\bibitem [{\citenamefont {Scully}\ and\ \citenamefont
  {Lamb~Jr}(1967)}]{scully1967quantum}%
  \BibitemOpen
  \bibfield  {author} {\bibinfo {author} {\bibfnamefont {M.~O.}\ \bibnamefont
  {Scully}}\ and\ \bibinfo {author} {\bibfnamefont {W.~E.}\ \bibnamefont
  {Lamb~Jr}},\ }\bibfield  {title} {\bibinfo {title} {Quantum theory of an
  optical maser. i. general theory},\ }\href@noop {} {\bibfield  {journal}
  {\bibinfo  {journal} {Physical Review}\ }\textbf {\bibinfo {volume} {159}},\
  \bibinfo {pages} {208} (\bibinfo {year} {1967})}\BibitemShut {NoStop}%
\bibitem [{\citenamefont {Scully}(1999)}]{scully1999condensation}%
  \BibitemOpen
  \bibfield  {author} {\bibinfo {author} {\bibfnamefont {M.~O.}\ \bibnamefont
  {Scully}},\ }\bibfield  {title} {\bibinfo {title} {Condensation of n bosons
  and the laser phase transition analogy},\ }\href@noop {} {\bibfield
  {journal} {\bibinfo  {journal} {Physical review letters}\ }\textbf {\bibinfo
  {volume} {82}},\ \bibinfo {pages} {3927} (\bibinfo {year}
  {1999})}\BibitemShut {NoStop}%
\bibitem [{\citenamefont {DeGiorgio}\ and\ \citenamefont
  {Scully}(1970)}]{degiorgio1970analogy}%
  \BibitemOpen
  \bibfield  {author} {\bibinfo {author} {\bibfnamefont {V.}~\bibnamefont
  {DeGiorgio}}\ and\ \bibinfo {author} {\bibfnamefont {M.~O.}\ \bibnamefont
  {Scully}},\ }\bibfield  {title} {\bibinfo {title} {Analogy between the laser
  threshold region and a second-order phase transition},\ }\href@noop {}
  {\bibfield  {journal} {\bibinfo  {journal} {Physical Review A}\ }\textbf
  {\bibinfo {volume} {2}},\ \bibinfo {pages} {1170} (\bibinfo {year}
  {1970})}\BibitemShut {NoStop}%
\bibitem [{\citenamefont {Graham}\ and\ \citenamefont
  {Haken}(1970)}]{graham1970laserlight}%
  \BibitemOpen
  \bibfield  {author} {\bibinfo {author} {\bibfnamefont {R.}~\bibnamefont
  {Graham}}\ and\ \bibinfo {author} {\bibfnamefont {H.}~\bibnamefont {Haken}},\
  }\bibfield  {title} {\bibinfo {title} {Laserlight—first example of a
  second-order phase transition far away from thermal equilibrium},\
  }\href@noop {} {\bibfield  {journal} {\bibinfo  {journal} {Zeitschrift
  f{\"u}r Physik}\ }\textbf {\bibinfo {volume} {237}},\ \bibinfo {pages} {31}
  (\bibinfo {year} {1970})}\BibitemShut {NoStop}%
\bibitem [{\citenamefont {Zhang}\ \emph {et~al.}(2021)\citenamefont {Zhang},
  \citenamefont {Wang},\ and\ \citenamefont {Wang}}]{zhang2021quantum}%
  \BibitemOpen
  \bibfield  {author} {\bibinfo {author} {\bibfnamefont {Z.}~\bibnamefont
  {Zhang}}, \bibinfo {author} {\bibfnamefont {X.}~\bibnamefont {Wang}},\ and\
  \bibinfo {author} {\bibfnamefont {J.}~\bibnamefont {Wang}},\ }\bibfield
  {title} {\bibinfo {title} {Quantum fluctuation-dissipation theorem far from
  equilibrium},\ }\href@noop {} {\bibfield  {journal} {\bibinfo  {journal}
  {Physical Review B}\ }\textbf {\bibinfo {volume} {104}},\ \bibinfo {pages}
  {085439} (\bibinfo {year} {2021})}\BibitemShut {NoStop}%
\bibitem [{\citenamefont {Jauslin}\ and\ \citenamefont
  {Sugny}(2010)}]{jauslin2010dynamics}%
  \BibitemOpen
  \bibfield  {author} {\bibinfo {author} {\bibfnamefont {H.-R.}\ \bibnamefont
  {Jauslin}}\ and\ \bibinfo {author} {\bibfnamefont {D.}~\bibnamefont
  {Sugny}},\ }\bibfield  {title} {\bibinfo {title} {Dynamics of mixed
  classical-quantum systems, geometric quantization and coherent states},\ }in\
  \href@noop {} {\emph {\bibinfo {booktitle} {Mathematical horizons for quantum
  physics}}}\ (\bibinfo  {publisher} {World Scientific},\ \bibinfo {year}
  {2010})\ pp.\ \bibinfo {pages} {65--96}\BibitemShut {NoStop}%
\bibitem [{\citenamefont {Bolterauer}(1999)}]{bolterauer1999elementary}%
  \BibitemOpen
  \bibfield  {author} {\bibinfo {author} {\bibfnamefont {H.}~\bibnamefont
  {Bolterauer}},\ }\bibfield  {title} {\bibinfo {title} {Elementary arguments
  that the wu--austin hamiltonian has no finite ground state (the search for a
  microscopic foundation of fr{\"o}hlichs theory)},\ }\href@noop {} {\bibfield
  {journal} {\bibinfo  {journal} {Bioelectrochemistry and bioenergetics}\
  }\textbf {\bibinfo {volume} {48}},\ \bibinfo {pages} {301} (\bibinfo {year}
  {1999})}\BibitemShut {NoStop}%
\bibitem [{\citenamefont {Scully}\ and\ \citenamefont
  {Zubairy}(1999)}]{scully1999quantum}%
  \BibitemOpen
  \bibfield  {author} {\bibinfo {author} {\bibfnamefont {M.~O.}\ \bibnamefont
  {Scully}}\ and\ \bibinfo {author} {\bibfnamefont {M.~S.}\ \bibnamefont
  {Zubairy}},\ }\href@noop {} {\bibinfo {title} {Quantum optics}} (\bibinfo
  {year} {1999})\BibitemShut {NoStop}%
\bibitem [{\citenamefont {Breuer}\ \emph {et~al.}(2002)\citenamefont {Breuer},
  \citenamefont {Petruccione} \emph {et~al.}}]{breuer2002theory}%
  \BibitemOpen
  \bibfield  {author} {\bibinfo {author} {\bibfnamefont {H.-P.}\ \bibnamefont
  {Breuer}}, \bibinfo {author} {\bibfnamefont {F.}~\bibnamefont {Petruccione}},
  \emph {et~al.},\ }\href@noop {} {\emph {\bibinfo {title} {The theory of open
  quantum systems}}}\ (\bibinfo  {publisher} {Oxford University Press on
  Demand},\ \bibinfo {year} {2002})\BibitemShut {NoStop}%
\bibitem [{\citenamefont {Agarwal}(2012)}]{agarwal2012quantum}%
  \BibitemOpen
  \bibfield  {author} {\bibinfo {author} {\bibfnamefont {G.~S.}\ \bibnamefont
  {Agarwal}},\ }\href@noop {} {\emph {\bibinfo {title} {Quantum optics}}}\
  (\bibinfo  {publisher} {Cambridge University Press},\ \bibinfo {year}
  {2012})\BibitemShut {NoStop}%
\bibitem [{\citenamefont {Wang}\ and\ \citenamefont
  {Wang}(2019)}]{wang2019nonequilibrium}%
  \BibitemOpen
  \bibfield  {author} {\bibinfo {author} {\bibfnamefont {X.}~\bibnamefont
  {Wang}}\ and\ \bibinfo {author} {\bibfnamefont {J.}~\bibnamefont {Wang}},\
  }\bibfield  {title} {\bibinfo {title} {Nonequilibrium effects on quantum
  correlations: Discord, mutual information, and entanglement of a
  two-fermionic system in bosonic and fermionic environments},\ }\href@noop {}
  {\bibfield  {journal} {\bibinfo  {journal} {Physical Review A}\ }\textbf
  {\bibinfo {volume} {100}},\ \bibinfo {pages} {052331} (\bibinfo {year}
  {2019})}\BibitemShut {NoStop}%
\bibitem [{\citenamefont {Wang}\ and\ \citenamefont
  {Wang}(2022)}]{wang2022effect}%
  \BibitemOpen
  \bibfield  {author} {\bibinfo {author} {\bibfnamefont {X.}~\bibnamefont
  {Wang}}\ and\ \bibinfo {author} {\bibfnamefont {J.}~\bibnamefont {Wang}},\
  }\bibfield  {title} {\bibinfo {title} {The effect of nonequilibrium entropy
  production on the quantum fisher information and correlations},\ }\href@noop
  {} {\bibfield  {journal} {\bibinfo  {journal} {Quantum Information
  Processing}\ }\textbf {\bibinfo {volume} {21}},\ \bibinfo {pages} {1}
  (\bibinfo {year} {2022})}\BibitemShut {NoStop}%
\bibitem [{\citenamefont {Mandel}(1982)}]{mandel1982squeezed}%
  \BibitemOpen
  \bibfield  {author} {\bibinfo {author} {\bibfnamefont {L.}~\bibnamefont
  {Mandel}},\ }\bibfield  {title} {\bibinfo {title} {Squeezed states and
  sub-poissonian photon statistics},\ }\href@noop {} {\bibfield  {journal}
  {\bibinfo  {journal} {Physical Review Letters}\ }\textbf {\bibinfo {volume}
  {49}},\ \bibinfo {pages} {136} (\bibinfo {year} {1982})}\BibitemShut
  {NoStop}%
\bibitem [{\citenamefont {Davidovich}(1996)}]{davidovich1996sub}%
  \BibitemOpen
  \bibfield  {author} {\bibinfo {author} {\bibfnamefont {L.}~\bibnamefont
  {Davidovich}},\ }\bibfield  {title} {\bibinfo {title} {Sub-poissonian
  processes in quantum optics},\ }\href@noop {} {\bibfield  {journal} {\bibinfo
   {journal} {Reviews of Modern Physics}\ }\textbf {\bibinfo {volume} {68}},\
  \bibinfo {pages} {127} (\bibinfo {year} {1996})}\BibitemShut {NoStop}%
\bibitem [{\citenamefont {Penrose}\ and\ \citenamefont
  {Onsager}(1956)}]{penrose1956bose}%
  \BibitemOpen
  \bibfield  {author} {\bibinfo {author} {\bibfnamefont {O.}~\bibnamefont
  {Penrose}}\ and\ \bibinfo {author} {\bibfnamefont {L.}~\bibnamefont
  {Onsager}},\ }\bibfield  {title} {\bibinfo {title} {Bose-einstein
  condensation and liquid helium},\ }\href@noop {} {\bibfield  {journal}
  {\bibinfo  {journal} {Physical Review}\ }\textbf {\bibinfo {volume} {104}},\
  \bibinfo {pages} {576} (\bibinfo {year} {1956})}\BibitemShut {NoStop}%
\bibitem [{\citenamefont {Yang}(1962)}]{yang1962concept}%
  \BibitemOpen
  \bibfield  {author} {\bibinfo {author} {\bibfnamefont {C.~N.}\ \bibnamefont
  {Yang}},\ }\bibfield  {title} {\bibinfo {title} {Concept of off-diagonal
  long-range order and the quantum phases of liquid he and of
  superconductors},\ }\href@noop {} {\bibfield  {journal} {\bibinfo  {journal}
  {Reviews of Modern Physics}\ }\textbf {\bibinfo {volume} {34}},\ \bibinfo
  {pages} {694} (\bibinfo {year} {1962})}\BibitemShut {NoStop}%
\end{thebibliography}%
\newpage
%\begin{widetext}
% \include{SM:phase_transition.tex}
%\end{widetext}
%%%%%%%%%%%%%%%%%%%%%%%
%%%%%%%%%%%%%%%%%%%%%%%
%%%%%%%%%%%%%%%%%%%%%%%

\begin{widetext}
\section{Supplementary Material}
In this Supplemental Material, we provide a detailed description of the methods used in the primary text as well as the detailed derivations of the important results.
\section{Model and rate equations}
We start from the Wu-Austin Hamiltonian as a model for nonequilibrium systems such as the collective oscillations of biomacromolecules in the solvent:
\begin{align}
    \mathcal{H}=&\sum_{i} \hbar \omega_{i} a_i^\dag a_i + \sum_{i} \hbar \bar\Omega_{i} b_i^\dag b_i + \sum_{i} \hbar \Omega_{i} p_i^\dag p_i +\hbar\sum_i\sum_k f_{i,k} a_i^{\dagger}b_k  
 + \hbar\sum_{i,j}\sum_k g_{ij,k}a_i^{\dagger}a_j b_k 
+\hbar\sum_i\sum_k \lambda_{i,k} a_i^{\dagger}p_k  + \text{h.c.} \,,
\label{sm:hamiltonian}
\end{align}
where $a_i$ is the annihilation operator for the vibrational modes, $b_i$ is the annihilation operator for the surrounding biological system (solvent), and $p_i$ is for the external source. In most biological settings, the solvent contains a large  number of degrees of freedom with a short decoherence time and can be modeled as the Markovian bath. With this Hamiltonian, we can write the equation of motion for the density matrix of the system using the standard techniques:
\begin{equation}
\begin{split}
\dot{\rho}_s = -\frac{1}{\hbar^2} \int_0^{\infty}\text{d}t'\ \text{Tr}_{B} \left[V_{\text{int}}(t),[V_{\text{int}}(t-t'),\rho_B\otimes\rho_s(t)]\right]
\end{split}
\end{equation}
where $\rho_s$ is the reduced density matrix of the system, $\rho_{B}$ is the density matrix of the bath and the external source, and $V_{\text{int}}$ is the interaction Hamiltonian in the interaction picture given by 
\begin{align}
    V_{\text{int}}=\hbar\sum_i\sum_k f_{i,k} a_i^{\dagger}b_k e^{i(\omega_i-\bar{\Omega}_k)t}  
 + \hbar\sum_{i,j}\sum_k g_{ij,k}a_i^{\dagger}a_j b_k e^{i(\omega_i-\omega_j-\bar{\Omega}_k)t} 
+\hbar\sum_i\sum_k \lambda_{i,k} a_i^{\dagger}p_k e^{i(\omega_i-\Omega_k)t} + \text{h.c.}\,.
\end{align}
Since we will be dealing with only the system density matrix from now on and we omit the subscript ``s" from the reduced density matrix $\rho_s$. After some algebra, we obtain the equations of motion for the system density operator: 
\begin{equation}
\begin{split}
\dot{\rho} =  &\frac{\phi}{2}\sum_{s=0}^D\left[(\bar{n}_{\omega_s}+1)\left(a_s\rho a_s^{\dagger}-a_s^{\dagger}a_s\rho\right)+\bar{n}_{\omega_s}\left(a_s^{\dagger}\rho a_s-a_s a_s^{\dagger}\rho\right)\right]\\[0.15cm]
&+\frac{\Lambda}{2}\sum_{s=0}^D\left[(n_{\omega_s}+1)\left(a_s\rho a_s^{\dagger}-a_s^{\dagger}a_s\rho\right)+n_{\omega_s}\left(a_s^{\dagger}\rho a_s-a_s a_s^{\dagger}\rho\right)\right]\\[0.15cm]
& + \frac{\chi}{2}\sum_{s=1}^D\sum_{j=0}^{s-1}\left[(\bar{n}_{\omega_{sj}}+1)\left(a_j^{\dagger}a_s\rho a_s^{\dagger}a_j-a_s^{\dagger}a_j a_j^{\dagger}a_s\rho\right)+ \bar{n}_{\omega_{sj}}\left(a_s^{\dagger}a_j\rho a_j^{\dagger}a_s-a_j^{\dagger}a_s a_s^{\dagger}a_j\rho\right)\right]  +\text{h.c.}\,,
\end{split}
\label{sm:qme}
\end{equation}
where $\bar{n}_{\omega}=[\text{exp}(\hbar\omega/k_B T)-1]^{-1}$ is the occupation number of the bath at frequency $\omega$, and $\phi=2\pi f_{\omega}^2{\cal D}(\omega),\ \Lambda=2\pi \lambda_{\omega}^{2}{\cal D}(\omega),\ \chi=2\pi g_{\omega}^2{\cal D}(\omega)$ are the rates of pumping, dissipation (one-phonon) and energy redistribution (two-phonon), respectively. ${\cal D}(\omega)$ is the density of states of the bath. Since the bandwidth of the oscillations $ \mathcal{I}_{sys}$ is narrow, the rates are approximated to be the same for each mode. To concentrate on the collective behavior of condensation, we further reduce the system density matrix to the $l^{\mathrm{th}}$ mode:
\begin{align}
    \rho_{l}:=\sum_{\{n_k\}}\langle \{n_k\}|\rho|\{n_k\}\rangle\,,\quad \{n_k\} = \{n_0,n_1,\cdots,n_{l-1},n_{l+1},\cdots,n_D\}\,,
\end{align}
where $\{n_k\}$ is the configuration of the phonons except the mode $l$. Hereafter, we further omit the subscripts for convenience. Then the mean phonon number at mode $\omega_l$ is $ \langle n_l \rangle  =\sum_{n_l=0}^\infty n_l \ \rho_{n_l,n_l}$.
The rate equation of phonon numbers on mode $l$ reads:
\begin{equation}
    \langle \dot{n}_l\rangle  =\sum_{n_l=0}^\infty n_l \ \dot\rho_{n_l,n_l}\,,
\end{equation}
where 
\begin{equation}
    \begin{split}
        \dot{\rho}_{n_l,n_l} = & \sum_{\{n_k\}}\langle n_l,\{n_k\}|\dot \rho|n_l,\{n_k\}\rangle\,,\\
        = & \phi \sum_{s=0}^D\sum_{\{n_k\}}\langle n_l,\{n_k\}|\left[(\bar{n}_{\omega_s}+1)\left(a_s\rho a_s^{\dagger}-a_s^{\dagger}a_s\rho\right)+\bar{n}_{\omega_s}\left(a_s^{\dagger}\rho a_s-a_s a_s^{\dagger}\rho\right)\right]|n_l,\{n_k\}\rangle\\[0.15cm]
        &+ \Lambda \sum_{s=0}^D\sum_{\{n_k\}}\langle n_l,\{n_k\}|\left[(\bar{n}_{\omega_s}+1)\left(a_s\rho a_s^{\dagger}-a_s^{\dagger}a_s\rho\right)+\bar{n}_{\omega_s}\left(a_s^{\dagger}\rho a_s-a_s a_s^{\dagger}\rho\right)\right]|n_l,\{n_k\}\rangle\\[0.15cm]
& +\chi \sum_{s=1}^D\sum_{j=0}^{s-1}\sum_{\{n_k\}}\langle n_l,\{n_k\}|\left[(\bar{n}_{\omega_{sj}}+1)\left(a_j^{\dagger}a_s\rho a_s^{\dagger}a_j-a_s^{\dagger}a_j a_j^{\dagger}a_s\rho\right) + \bar{n}_{\omega_{sj}}\left(a_s^{\dagger}a_j\rho a_j^{\dagger}a_s-a_j^{\dagger}a_s a_s^{\dagger}a_j\rho\right)\right]|n_l,\{n_k\}\rangle \,.
    \end{split}
    \label{sm:rhodot}
\end{equation}
The calculation of the above rate equation can be greatly simplified if we decompose the triple summations proportional to $\chi$ by the following rule:
\begin{equation}
\begin{split}
      \sum_{s=1}^D\sum_{j=0}^{s-1}\sum_{\{n_k\}} =&\sum_{s<l}\sum_{j=0}^{s-1}\sum_{\{n_k\}}+\sum_{s=l}^l\sum_{j=0}^{l-1}\sum_{\{n_k\}}+\sum_{s>l}\sum_{j=0}^{s-1}\sum_{\{n_k\}} \\
      =& \sum_{s<l}\sum_{j=0}^{s-1}\sum_{\{n_k\}}+\sum_{s=l}^l\sum_{j=0}^{l-1}\sum_{\{n_k\}}+\Big( \sum_{s>l}^D\sum_{j\ne l}^{s-1}\sum_{\{n_k\}}+\sum_{s>l}^D\sum_{j=l}^l\sum_{\{n_k\}}\Big)\\
      =&\sum_{s<l}\sum_{j=0}^{s-1}\sum_{\{n_k\}}+ \sum_{s>l}^D\sum_{j\ne l}^{s-1}\sum_{\{n_k\}}+\sum_{s>l}^D\sum_{j=l}^l\sum_{\{n_k\}}+\sum_{s=l}^l\sum_{j=0}^{l-1}\sum_{\{n_k\}}\,.
      \label{sm:summations}
\end{split}
\end{equation}
It is easy to check that only the last two summations in the last line of Eq.~\eqref{sm:summations} contribute to the $\dot\rho_{n_l,n_l}$ and the first two summations are identically zero. After some manipulation of algebra, we find the expression of Eq.~\eqref{sm:rhodot} to be the following:
\begin{equation}
\begin{split}
\dot{\rho}_{n_l,n_l} = & \left(\phi(\bar{n}+1) +\Lambda (n_{ex}+1)\right)\Big((n_l+1)\rho_{n_l+1,n_l+1} - n_l\rho_{n_l,n_l}\Big) + \left(\phi\bar{n}+\Lambda n_{ex} \right)\Big(n_l\rho_{n_l-1,n_l-1} - (n_l+1)\rho_{n_l,n_l}\Big)\\[0.15cm]
& + \chi\sum_{s>l}^D\bigg[(\bar{n}_{\omega_{sl}}+1)\sum_{\{n_j\}}\Big(n_l n_s\langle n_l-1;\{n_j\}|\rho|n_l-1;\{n_j\}\rangle - (n_l+1)n_s\langle n_l;\{n_j\}|\rho|n_l;\{n_j\}\rangle\Big)\\[0.15cm]
&  + \bar{n}_{\omega_{sl}}\sum_{\{n_j\}}\Big((n_l+1)(n_s+1)\langle n_l+1;\{n_j\}|\rho|n_l+1;\{n_j\}\rangle - n_l(n_s+1)\langle n_l;\{n_j\}|\rho|n_l;\{n_j\}\rangle\Big)\bigg]\\[0.15cm]%%%%%%%%%%
& + \chi\sum_{j<l}\bigg[(\bar{n}_{\omega_{lj}}+1)\sum_{\{n_j\}}\Big((n_l+1)(n_j+1)\langle n_l+1;\{n_j\}|\rho|n_l+1;\{n_j\}\rangle - n_l(n_j+1)\langle n_l;\{n_j\}|\rho|n_l;\{n_j\}\rangle\Big)\\[0.15cm]
&  + \bar{n}_{\omega_{lj}}\sum_{\{n_j\}}\Big(n_l n_j\langle n_l-1;\{n_j\}|\rho|n_l-1;\{n_j\}\rangle - (n_l+1)n_j \langle n_l;\{n_j\}|\rho|n_l;\{n_j\}\rangle\Big)\bigg]\,,
\end{split}
\label{sm:rhondot}
\end{equation}
where we have dropped the indices on the bath particle number $\bar{n}_s$ for notational convenience. Using the dynamics of the state Eq.~\eqref{sm:rhondot}, we obtain the rate equation as the following:
\begin{equation}
\begin{split}
\langle \dot{n}_l\rangle =&  \left(\phi(\bar{n}+1)+\Lambda (n_{ex}+1)\right)\sum_{n_l=0}^\infty n_l\Big((n_l+1)\rho_{n_l+1,n_l+1} - n_l\rho_{n_l,n_l}\Big) + \left(\phi\bar{n}+\Lambda n_{ex}\right)\sum_{n_l=0}^\infty n_l\Big(n_l\rho_{n_l-1,n_l-1} - (n_l+1)\rho_{n_l,n_l}\Big)\\[0.15cm]
& + \chi\sum_{s>l}^D\bigg[(\bar{n}_{\omega_{sl}}+1)\sum_{\{n_j\}}\sum_{n_l=0}^\infty n_l\Big(n_l n_s\langle n_l-1;\{n_j\}|\rho|n_l-1;\{n_j\}\rangle - (n_l+1)n_s\langle n_l;\{n_j\}|\rho|n_l;\{n_j\}\rangle\Big)\\[0.15cm]
&  + \bar{n}_{\omega_{sl}}\sum_{\{n_j\}}\sum_{n_l=0}^\infty n_l\Big((n_l+1)(n_s+1)\langle n_l+1;\{n_j\}|\rho|n_l+1;\{n_j\}\rangle - n_l(n_s+1)\langle n_l;\{n_j\}|\rho|n_l;\{n_j\}\rangle\Big)\bigg]\\%%
& + \chi\sum_{j<l}\bigg[(\bar{n}_{\omega_{lj}}+1)\sum_{\{n_j\}}\sum_{n_l=0}^\infty n_l\Big((n_l+1)(n_j+1)\langle n_l+1;\{n_j\}|\rho|n_l+1;\{n_j\}\rangle - n_l(n_j+1)\langle n_l;\{n_j\}|\rho|n_l;\{n_j\}\rangle\Big)\\[0.15cm]
&  + \bar{n}_{\omega_{lj}}\sum_{\{n_j\}}\sum_{n_l=0}^\infty n_l\Big(n_l n_j\langle n_l-1;\{n_j\}|\rho|n_l-1;\{n_j\}\rangle - (n_l+1)n_j \langle n_l;\{n_j\}|\rho|n_l;\{n_j\}\rangle\Big)\bigg]\,.\\[0.15cm]%%%%%%%%%%
\end{split}
\end{equation}
To simplify the rate equation $\langle \dot n_l\rangle$, we apply the following identity:
\begin{equation}
   \sum_{n_l=0}^\infty n_l \ \dot\rho_{n_l,n_l}=\sum_{n_l=1}^\infty (n_l-1) \ \dot\rho_{n_l-1,n_l-1}=\sum_{n_l=0}^\infty (n_l+1) \ \dot\rho_{n_l+1,n_l+1}\,,
\end{equation}
and rewrite the rate equation as follows:
\begin{equation}
\begin{split}
\langle \dot{n}_l\rangle =& \left(\phi(\bar{n}+1)+\Lambda (n_{ex}+1)\right)\sum_{n_l=0}^\infty \Big(\big((n_l+1)^2-(n_l+1)\big)\rho_{n_l+1,n_l+1} - n_l^2\rho_{n_l,n_l}\Big) \\
&+ \left(\phi\bar{n}+\Lambda n_{ex}\right)\sum_{n_l=0}^\infty \Big(\big((n_l-1)+1\big)^2\rho_{n_l-1,n_l-1} - (n_l^2+n_l)\rho_{n_l,n_l}\Big)\\[0.15cm]
& + \chi\sum_{s>l}^D\bigg[(\bar{n}_{\omega_{sl}}+1)\sum_{\{n_j\}}\sum_{n_l=0}^\infty \Big(\big((n_l-1)+1\big)^2 n_s\langle n_l-1;\{n_j\}|\rho|n_l-1;\{n_j\}\rangle - (n_l^2+n_l)n_s\langle n_l;\{n_j\}|\rho|n_l;\{n_j\}\rangle\Big)\\[0.15cm]
&  + \bar{n}_{\omega_{sl}}\sum_{\{n_j\}}\sum_{n_l=0}^\infty \Big((n_l+1)(n_l+1-1)(n_s+1)\langle n_l+1;\{n_j\}|\rho|n_l+1;\{n_j\}\rangle - n_l^2(n_s+1)\langle n_l;\{n_j\}|\rho|n_l;\{n_j\}\rangle\Big)\bigg] \\[0.15cm]
& + \chi\sum_{j<l}\bigg[(\bar{n}_{\omega_{lj}}+1)\sum_{\{n_j\}}\sum_{n_l=0}^\infty \Big((n_l+1)(n_l+1-1)(n_j+1)\langle n_l+1;\{n_j\}|\rho|n_l+1;\{n_j\}\rangle - n_l^2(n_j+1)\langle n_l;\{n_j\}|\rho|n_l;\{n_j\}\rangle\Big)\\[0.15cm]
&  + \bar{n}_{\omega_{lj}}\sum_{\{n_j\}}\sum_{n_l=1}^\infty \Big((n_l+1)^2 n_j\langle n_l-1;\{n_j\}|\rho|n_l-1;\{n_j\}\rangle - (n_l^2+n_l)n_j \langle n_l;\{n_j\}|\rho|n_l;\{n_j\}\rangle\Big)\bigg]\\%%%%%%%%
\end{split}
\end{equation}
The summations now can be easily computed and finally we have the rate equation: 
\begin{equation}
\begin{split}
\langle \dot{n}_l\rangle =& \left(\phi(\bar{n}+1)+\Lambda (n_{ex}+1)\right) \Big(\langle n_l^2-n_l \rangle - \langle n_l^2 \rangle\Big) + \left(\phi\bar{n}+\Lambda n_{ex}\right) \Big(\langle n_l^2+2n_l+1\rangle-\langle n_l^2+n_l \rangle \Big)\\[0.15cm]
& + \chi\sum_{s>}^D\bigg[(\bar{n}_{\omega_{sl}}+1) \Big(\langle n_s(n_l^2+2n_l+1) \rangle - \langle n_s(n_l^2+n_l)\rangle\Big)  + \bar{n}_{\omega_{sl}} \Big(\langle (n_s+1)n_l^2-n_l(n_s+1)\rangle - \langle n_l^2(n_s+1)\rangle\Big)\bigg]\\
&+  \chi\sum_{j<l}\bigg[(\bar{n}_{\omega_{sl}}+1) \Big(\langle (n_j+1)(n_l^2-n_l) \rangle - \langle (n_j+1)n_l^2\rangle\Big) + \bar{n}_{\omega_{sl}} \Big(\langle n_j(n_l+1)^2\rangle - \langle (n_l^2+n_l)n_j)\rangle\Big)\bigg] \\[0.15cm]%%%%%%%%%%%%
=& \ (\phi\bar{n}_{\omega_l}+\Lambda n_{ex}) \langle n_l+1\rangle -\left(\phi\,(\bar{n}_{\omega_l}+1)++\Lambda (n_{ex}+1)\right) \langle n_l \rangle 
 +\chi\sum_{s=l+1}^D\left[(\bar{n}_{\omega_{sl}}+1) \langle n_s(n_l+1) \rangle - \bar{n}_{\omega_{sl}} \langle (n_s+1)n_l \rangle \right] \\
 &\ \ + \chi\sum_{j<l}\left[-(\bar{n}_{\omega_{jl}}+1) \langle (n_j+1)n_l \rangle + \bar{n}_{\omega_{jl}} \langle n_j(n_l+1)\rangle \right]  \,.
\end{split}
\label{sm:rateeqn}
\end{equation}
To recapitulate, the above algebra gives the simple rate equation of phonon numbers at mode $\omega_l$ to be:
\begin{equation}
\begin{split}
\langle \dot{n}_l\rangle  =& 
\ \phi\big[\bar{n}_{\omega_l} - \langle n_l\rangle\big]  +  \Lambda \big[n_{ex} - \langle n_l\rangle\big] +\chi\Big\{\sum_{j>l}\big[\bar{n}_{\omega_{jl}}\langle n_j-n_l \rangle +\langle n_jn_l+n_j\rangle\big] + \sum_{j<l}\big[\bar{n}_{\omega_{lj}}\langle n_j-n_l\rangle - \langle n_j n_l+n_l\rangle\big]\Big\}\,.
\end{split}
\end{equation}
Given the above equation, the rate equation of the total number of phonons $N=\sum_{s=0}^D a_s^{\dagger}a_s$ can be calculated under the approximation of the smooth spectrum of the bath $\bar{n}_{\omega_l}\simeq \bar{n}_{\omega_0}\equiv\bar{n},\ \bar{n}_{\omega_{jl}}\simeq \bar{n}_{\omega_0}\equiv\bar{n}$:
\begin{equation}
\begin{split}
\langle\dot{N}\rangle  =& (D+1)(\phi \bar{n}+\Lambda n_{ex}) - (\phi+\Lambda)\langle N\rangle + \chi \Big(\sum_{l=0}^{D-1}\sum_{j=l+1}^D\langle n_j(n_l+1)\rangle - \sum_{l=1}^D\sum_{j=0}^{l-1}\langle (n_j+1)n_l\rangle\Big)\,.
\end{split}
\end{equation}
Here, the nonlinear term vanishes identically. Therefore, in the NESS the total number of phonons is: 
\begin{equation}
       \langle N\rangle\simeq \frac{(D+1)(\phi\bar{n}+\Lambda n_{ex})}{\phi+\Lambda}\,.
\end{equation} 
In particular, the rate equation for the condensate phonon number $n_0$ according to Eq.~\eqref{sm:rateeqn} is:
\begin{equation}
\begin{split}
\langle \dot{n}_0\rangle = & \phi\big[\bar{n}(\langle n_0\rangle+1) - (\bar{n}+1)\langle n_0\rangle\big]+\Lambda\big[n_{ex}(\langle n_0\rangle+1) - (n_{ex}+1)\langle n_0\rangle\big]\\
&+ \chi\sum_{s=1}^D\big[(\bar{n}_{\omega_{s0}}+1)\langle n_s\rangle(\langle n_0\rangle+1) - \bar{n}_{\omega_{s0}}(\langle n_s\rangle+1)\langle n_0\rangle\big]
\end{split}
\label{sm:n0}
\end{equation}
where in the $\chi$-part in Eq.~\eqref{sm:n0}, the first term describes the process that the mode $\omega_s$ decays to mode $\omega_0$ together with the dissipation of one quanta to the bath with frequency $\omega_s-\omega_0$. The reverse process is described by the second term. At the steady state, the phonon number at mode $\omega_0$ can be formally expressed as
\begin{align}
\langle n_0\rangle &= \frac{\phi\bar{n}+\chi(\bar{n}+1)\langle N_e\rangle+\Lambda n_{ex}}{\phi+\Lambda-\chi(\langle N_e\rangle-D\bar{n})}\\
&= \frac{\phi\bar{n}+\chi(\bar{n}+1)(\langle N\rangle-\langle n_0 \rangle)+\Lambda n_{ex}}{\phi+\Lambda-\chi(\langle N\rangle-\langle n_0 \rangle-D\bar{n})}\,.
\label{sm:n0sol}
\end{align}

The nonlinear terms proportional to $\chi$ is essential in the condensation process. It is easy to check that by setting $\chi=0$, the mean value of phonon number at mode $\omega_0$ is given by the thermal average of the environment and the external source:
\begin{equation}
    n_0=\dfrac{\phi \bar n+\Lambda n_{ex}}{\phi+\Lambda}\,.
\end{equation}
It reduces to the thermal distribution in the case of a single bath.

%%%%%%%%%%%%%%%%%%%%%
%%%%%%%%%%%%%%%%%%%%%
\section{Statistics and fluctuations}
To study the statistics and the fluctuations of the phonon condensate, we will only assume the decorrelation between the condensate state $n_0$ and total phonon number $\langle N\rangle$, $\langle n_0 N\rangle\simeq \langle N\rangle\langle n_0\rangle$, rather than the decorrelation approximation $\langle n_i n_j\rangle\simeq \langle n_i\rangle\langle n_j\rangle$ as used in Fr\"ohlich's original work. This approximation is due to the fact that the total phonon number $\langle N\rangle$ is solely dictated by pumping and dissipation (one phonon processes) instead of the nonlinear two-phonon process which correlates individual modes. Besides, the total phonon number is a large number with low relative fluctuations compared with each individual mode. We proceed by defining the further reduced density matrix of the condensate at the mode $\omega_0$:
\begin{equation}
\begin{split}
\rho_{n_0,m_0} = \sum_{\{n_l\}}\langle n_0,\{n_l\}|\rho|m_0,\{n_l\}\rangle,\quad \{n_l\} = \{n_1,n_2,\cdots,n_D\}\,.
\end{split}
\label{sm:rho0}
\end{equation}
Restrained only to the lowest vibrational state, Eq.~\eqref{sm:rhondot} gives
\begin{equation}
\begin{split}
\dot{\rho}_{n_0,n_0} = & \left(\phi(\bar{n}+1)\right)\Big((n_0+1)\rho_{n_0+1,n_0+1} - n_0\rho_{n_0,n_0}\Big) + \left(\phi\bar{n}\right)\Big(n_0\rho_{n_0-1,n_0-1} - (n_0+1)\rho_{n_0,n_0}\Big)\\[0.15cm]
& +\left(\Lambda(n_{ex}+1)\right)\Big((n_0+1)\rho_{n_0+1,n_0+1} - n_0\rho_{n_0,n_0}\Big) + \left(\Lambda n_{ex} \right)\Big(n_0\rho_{n_0-1,n_0-1} - (n_0+1)\rho_{n_0,n_0}\Big)\\[0.15cm]
& + \chi\sum_{s=1}^D \left[(\bar{n}_{\omega_{s0}}+1)\sum_{\{n_l\}}\Big(n_0 n_s\langle n_0-1;\{n_l\}|\rho|n_0-1;\{n_l\}\rangle - (n_0+1)n_s\langle n_0;\{n_l\}|\rho|n_0;\{n_l\}\rangle\Big) \right.\\[0.15cm]
& \quad \left.+ \bar{n}_{\omega_{s0}}\sum_{\{n_l\}}\Big((n_0+1)(n_s+1)\langle n_0+1;\{n_l\}|\rho|n_0+1;\{n_l\}\rangle - n_0(n_s+1)\langle n_0;\{n_l\}|\rho|n_0;\{n_l\}\rangle\Big)\right]
\end{split}
\label{sm:rhol}
\end{equation}
Using the smooth spectrum approximation, 
\begin{equation}
\begin{split}
\sum_{s=1}^D(\bar{n}_{\omega_{s0}}+1)\langle n_s\rangle_{n_0}=(\bar{n}+1)(\langle N\rangle-n_0)\,, \mathrm{and} \quad  \sum_{s=1}^D\bar{n}_{\omega_{s0}}\langle n_s+1\rangle_{n_0}=\bar{n}(\langle N\rangle-n_0+D)\,,
\end{split}
\end{equation}
one has the following equations
\begin{equation}
\begin{split}
\dot{\rho}_{n_0,n_0} = & - \left(\phi\bar{n}+\Lambda n_{ex}+\chi(\bar{n}+1)(\langle N\rangle-n_0)\right)(n_0+1)\rho_{n_0,n_0} \\
&+ \left(\phi\bar{n}+\Lambda n_{ex}+\chi(\bar{n}+1)(\langle N\rangle-n_0+1)\right)n_0\rho_{n_0-1,n_0-1}\\[0.15cm]
& - \left(\phi(\bar{n}+1)+\Lambda(n_{ex}+1)+\chi\bar{n}(\langle N\rangle-n_0+D)\right)n_0\rho_{n_0,n_0} \\
&+ \left(\phi(\bar{n}+1)+\Lambda(n_{ex}+1)+\chi\bar{n}(\langle N\rangle-n_0-1+D)\right)(n_0+1)\rho_{n_0+1,n_0+1}\,.
\end{split}
\label{sm:rhon0dot}
\end{equation}

Now that we have reduced the density matrix into the lowest diagonal mode $\rho_{n_0,n_0} = \sum_{\{n_l\}}\langle n_0,\{n_l\}|\rho|n_0,\{n_l\}\rangle$, where $\{n_l\} = \{n_1,n_2,\cdots,n_D\}$ is the configuration of the excited states and $n_0$ is the variable denoting the phonon number in the condensate. For any specific $n_0$, the diagonal element has the physical interpretation of the probability of finding $n_0$ phonons on the lowest mode. For simplicity, we denote the probability as $P(m)$ where $m \in \{0,1,2,3...\}$. To solve the phonon distribution function $P(n_0)$ at the nonequilibrium steady state, notice that we can regroup Eq.~\eqref{sm:rhon0dot} into two equivalent equations:
\begin{eqnarray}
\left(\phi\bar{n}+\Lambda n_{ex}+\chi(\bar{n}+1)\left(\langle N\rangle-n_0\right)\right)(n_0+1)\rho_{n_0,n_0} &-&    \left(\phi(\bar{n}+1)+\Lambda(n_{ex}+1)+\chi\bar{n}\left(\langle N\rangle-n_0-1+D\right)\right)(n_0+1)\rho_{n_0+1,n_0+1}=0\,, \nonumber\\
\left(\phi\bar{n}+\Lambda n_{ex}+\chi(\bar{n}+1)\left(\langle N\rangle-n_0+1\right)\right)n_0\rho_{n_0-1,n_0-1} &-&  \left(\phi(\bar{n}+1)+\Lambda(n_{ex}+1)+\chi\bar{n}\left(\langle N\rangle-n_0+D\right)\right)n_0\,\rho_{n_0,n_0}\quad =0\,.
\end{eqnarray}
Notice that the series of equations of different $n_0$'s are now related by geometric progression whose solution for $m$ condensate phonons is:
\begin{align}
P(m) =P(0)\ \left(1+\dfrac{1}{\bar{n}}\right)^{m}\ \dfrac{\left(\langle N\rangle+1+\dfrac{\phi \bar{n}+\Lambda n_{e}}{\chi (\bar{n}+1)}-m\right)_{(m)}}{\left(\langle N\rangle+D+\dfrac{\phi (\bar{n}+1)+\Lambda (n_e+1)}{\chi \bar{n}}-m\right)_{(m)}}\,,
\end{align}
where $(q)_{(m)}$ is the rising Pochhammer symbol. Equivalently. it can be rewritten in terms of Gamma functions:
\begin{equation}
\begin{split}
P(m) =P(0)\ \left(1+\dfrac{1}{\bar{n}}\right)^{m}\ \dfrac{\Gamma\left(\langle N\rangle+1+\dfrac{\phi \bar{n}+\Lambda n_{e}}{\chi (\bar{n}+1)}\right) \Gamma\left(\langle N\rangle+D-m+\dfrac{\phi (\bar{n}+1)+\Lambda (n_e+1)}{\chi \bar{n}}\right)}{\Gamma\left(\langle N\rangle-m+1+\dfrac{\phi \bar{n}+\Lambda n_e}{\chi (\bar{n}+1)}\right) \Gamma\left(\langle N\rangle+D+\dfrac{\phi (\bar{n}+1)+\Lambda (n_e+1)}{\chi \bar{n}}\right)}\,.
\end{split}
\end{equation}
Then, we redefine variables as the following: 
\begin{align}
    \mathcal{N}&= \langle N\rangle+1+\dfrac{\phi \bar{n}+\Lambda n_{e}}{\chi (\bar{n}+1)}\equiv \langle N\rangle+\alpha=\frac{(D+1)(\phi\bar{n}+\Lambda n_{ex})}{\phi+\Lambda}+\alpha\\
    \mathcal{D}&=D+\dfrac{\phi (\bar{n}+1)+\Lambda (n_e+1)}{\chi \bar{n}}-\dfrac{\phi \bar{n}+\Lambda n_{e}}{\chi (\bar{n}+1)}-1\equiv D+\beta\,.
    \label{sm:nd}
\end{align}
After substitution of variables, the probability distribution can be rewritten simply as:
\begin{equation}
\begin{split}
P(m) &=P(0)\ \left(1+\dfrac{1}{\bar{n}}\right)^{m} \frac{\left(\mathcal{N}-m\right)_{(m)}}{(\mathcal{N}+\mathcal{D}-m)_{(m)}}   \\
&=P(0)\ \left(1+\dfrac{1}{\bar{n}}\right)^{m}\ \dfrac{\Gamma\left(\mathcal{N}\right) \Gamma\left(\mathcal{N}+\mathcal{D}-m\right)}{\Gamma\left(\mathcal{N}-m\right) \Gamma\left(\mathcal{N}+\mathcal{D}\right)}\,.
\label{sm:prob}
\end{split}
\end{equation}
The probability of zero condensate $P(0)$ can be determined by the normalization condition $\sum_m P(m)=1$. This summation is the power series expansion of the hypergeometric function given as follows:
\begin{align}
    P(0)^{-1}=\, _2F_1\left(1,-\mathcal{N};-\mathcal{D}-\mathcal{N}+1;1+\frac{1}{\bar{n}}\right)\,,
\end{align}
where $_2F_1\left(1,-\mathcal{N};-\mathcal{D}-\mathcal{N}+1;1+\dfrac{1}{\bar{n}}\right)$ is the Gauss hypergeometric function of type $(2,1)$. As a reminder, $D$ is the total number of oscillators in the system and $\langle N \rangle$ is the total number of phonons in all modes in the NESS which is a function of $D$. The distribution of the phonons on the condensate corresponding to Eq.~\eqref{sm:prob} is shown in Fig.~\ref{Fig:sm1} of the Supplementary Material. 

The mean phonon numbers in the condensate can be calculated by the summation $\langle n_0\rangle=\sum_m mP(m)$, and we obtain:
\begin{align}
   \langle n_0\rangle=\frac{\mathcal{N} (1+\frac{1}{\bar{n}}) \, _2F_1(2,1-\mathcal{N};-\mathcal{D}-\mathcal{N}+2;1+\frac{1}{\bar{n}})}{ (\mathcal{D}+\mathcal{N}-1) \, _2F_1\left(1,-\mathcal{N};-\mathcal{D}-\mathcal{N}+1;1+\frac{1}{\bar{n}}\right)}\,.
   \label{sm:fra}
\end{align}
Similarly, we can calculate the mean of the second moment of the condensate phonon numbers. The summation returns:
\begin{align}
    \langle n_0^2\rangle =\frac{\mathcal{N} (1+\frac{1}{\bar{n}}) \, _3F_2(2,2,1-\mathcal{N};1,-\mathcal{D}-\mathcal{N}+2;1+\frac{1}{\bar{n}})}{(\mathcal{D}+\mathcal{N}-1)\,_2F_1\left(1,-\mathcal{N};-\mathcal{D}-\mathcal{N}+1;1+\frac{1}{\bar{n}}\right)}\,.
\end{align}
The Mandel Q parameter $Q=({\langle n_0^2\rangle -\langle n_0 \rangle^2})/{\langle n_0 \rangle} -1$ can be directly computed thereafter.

\section{Phase transition}
The phase behavior of the Fr\"ohlich condensation can be revealed by taking the large-$D$, or equivalently the large-$\langle N\rangle$ limit. The hypergeometric function given in Eq.~\eqref{sm:fra} is continuous for finite $\langle N\rangle$, which is expected from the finite size quantum system. In this case, there is no clear line or qualitative distinction between different regimes in the parameter space or ``phases". In this section, we will examine its behavior in the $D\rightarrow\infty$ limit.

 \begin{figure}[t]
    \centering
    \includegraphics[width=0.55\textwidth]{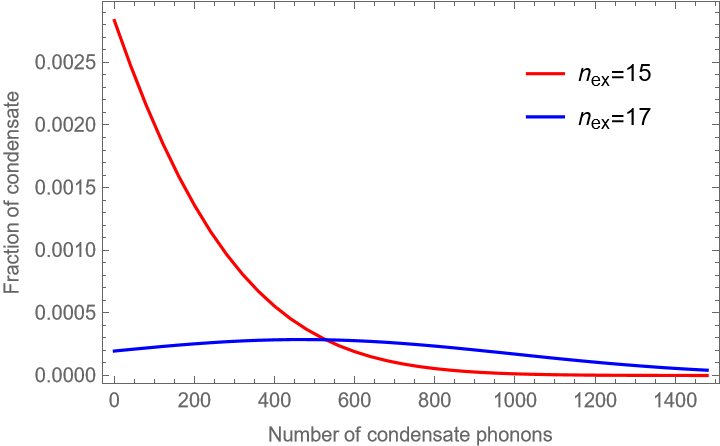}
    \caption{Distribution of condensate phonons. The red curve that monotonically decreases is for $n_{ex}=15<\bar n$ (below the transition point), and the blue curve is for $n_{ex}=17<\bar n$ (above the transition point).  Parameters are set to $\bar{n}=16,\,\Lambda=10\mathrm{GHz},\,\phi=5\mathrm{GHz}$, $\chi=0.07\mathrm{GHz}$ and $D=1000$.}
    \label{Fig:sm1}
\end{figure}

To study the fraction of phonons in the condensate, the crucial quantity to investigate is the probability distribution $P(m)$ in the case of large $m$. For example, we can look at the mean phonon number in the condensate $\langle n_0\rangle=\sum_{m}^{\mathcal{N}} mP(m)$. It is obvious that for $\langle n_0\rangle/\langle N\rangle$ to be nonzero as $D\rightarrow \infty$, the mean condensate phonon number has to scale as $\langle N\rangle$, namely, $\langle n_0\rangle\sim \langle N\rangle$. Therefore, the condition of no condensate $\langle n_0\rangle/\langle N\rangle\xrightarrow{D \rightarrow \infty}0$ is equivalent to $\langle N\rangle P(\langle N\rangle)\xrightarrow{D\rightarrow\infty}0$. This is further equivalent to the statement that there exists a positive integer $M\in N$ such that
\begin{align}
    \lim_{D\rightarrow\infty} P(m)/P(m-1)<1 \quad \mathrm{for\ all}\quad m>M\,.
\end{align}
Using Eq.~\eqref{sm:prob}, we have:
\begin{align}
    \frac{P(m)}{P(m-1)}&=\frac{\bar{n}+1}{\bar{n}}\cdot \frac{[\mathcal{N}-m]_{(m)}}{[\mathcal{N}+\mathcal{D}-m]_{(m)}}\cdot \frac{[\mathcal{N}+\mathcal{D}-(m-1)]_{(m-1)}}{[\mathcal{N}-(m-1)]_{(m-1)}}\\
    &=\frac{\bar{n}+1}{\bar{n}} \frac{\mathcal{N}-m}{\mathcal{N}+\mathcal{D}-m}\,,
\end{align}
where definitions of $\mathcal{N}$ and $\mathcal{D}$ are given in Eq.~\eqref{sm:nd}. Using the definitions of $\mathcal{N}$ and $\mathcal{D}$ and after some manipulation, we have the inequality from $P(m)/P(m-1)<1$ as:
\begin{align}
    \frac{(D+1)\Lambda}{\phi+\Lambda}(n_{ex}-\bar{n})<\frac{\phi+\Lambda}{\chi}-2 \bar{n}+m-1\,.
\end{align}
It is obvious that in the large-$D$ limit the condition of no condensate is that the external source occupation is below the occupation of the bath, i.e., $n_{ex}<\bar{n}$. Therefore, in the large-$D$ limit, 
\begin{align}
    \frac{\langle n_0\rangle}{\langle N\rangle}=0,\quad \mathrm{when} \quad n_{ex}<\bar{n}\,.
\end{align}
Furthermore, we simultaneously obtain the condition for condensation to emerge, which is $P(n_{cr}+1)>P(n_{cr})$ for some $n_{cr}\ge 1$. The inequality gives the condensation condition 
\begin{align}
    n_{ex}>\frac{(D-1)\Lambda-2\phi}{(D+1)\Lambda}\bar{n}+\left(\frac{\phi+\Lambda}{\chi}+n_{cr}-1\right)\frac{\phi+\Lambda}{(D+1)\Lambda}\,.
    \label{sm:n_critical}
\end{align}
Similarly, in the large D limit, the critical number density of the bath reduces to:
\begin{align}
     n_{ex}=\bar{n}\,.
\end{align}
This corresponds to the critical temperature $T_{cr}=T_1$. In other words, when the pump temperature is lower than that of the surrounding medium, no condensate will form. This is contrary to the situation of BEC where condensate appears only in the low temperature limit.

%The fraction of the condensate is solely determined by the asymptotic value of $\lim_{m\rightarrow\infty}mP(m)$ when such limit exists. 

In addition to the analysis given above, one can also directly study the hypergeometric functions. The hypergeometric functions have the expansion as the following:
\begin{align}
    _2F_1(a,-b;-c;d)=\sum_{m=0}^b \frac{b!}{m!(b-m)!}\frac{(a)_{(m)}}{(c)_{(m)}}d^m=\sum_{m=0}^b (m+1)_{(a-1)}\frac{(b-m+1)_{(m)}}{(c)_{(m)}}d^m \,,
\end{align}
where the subscript $(b)_{(m)}$ is the rising Pochhammer symbol defined by $(b)_{(m)}=b(b+1)...(b+m-1)$. Notice that $(m+1)_{(a-1)}$ is the polynomial of $m$ of order $(a-1)$. For $\frac{c}{b}>d$, we can assume without loss of generality that $\frac{bd}{c} \le k<1$. In this case. one can notice that 
\begin{align}
    \frac{(b-m+1)_{(m)}}{(c)_{(m)}}d^m<\frac{(b)^m}{(c)^m}d^m\le k^m\,.
\end{align}
Therefore, the summation
\begin{align}
    _2F_1(a,-b;-c;d)=\sum_{m}^b (m+1)_{(a-1)}\frac{(b-m+1)_{(m)}}{(c)_{(m)}}d^m< \sum_{m}^b (m+1)_{(a-1)}k^m
\end{align}
is convergent. The boundedness of such summation can be easily proved since there exists an $M\in \mathbf{N}$ such that for $m>M$,
\begin{align}
    (m+1)_{(a-1)}< (1/k)^{m/2}\,,
\end{align}
hence
\begin{align}
    \sum_{m}^b (m+1)_{(a-1)}k^m<\sum_{m}^b k^{m/2}\,.
\end{align}
This proves the finiteness of the hypergeometric functions in the case $\frac{c}{b}>d$, where $a,b,c \in \mathbf{N}$ and $d>0$. One may notice that in the limit $D\to\infty$,
\begin{align}
  \frac{\langle n_0\rangle}{\langle N \rangle}&=\frac{\mathcal{N} (1+\frac{1}{\bar{n}}) \, _2F_1(2,1-\mathcal{N};-\mathcal{D}-\mathcal{N}+2;1+\frac{1}{\bar{n}})}{ (\mathcal{D}+\mathcal{N}-1) \, _2F_1\left(1,-\mathcal{N};-\mathcal{D}-\mathcal{N}+1;1+\frac{1}{\bar{n}}\right)}/\langle N\rangle\\
  &\to  \frac{(1+\frac{1}{\bar{n}})}{(\mathcal{D}+\mathcal{N})}\frac{ _2F_1(2,1-\mathcal{N};-\mathcal{D}-\mathcal{N}+2;1+\frac{1}{\bar{n}})}{ _2F_1\left(1,-\mathcal{N};-\mathcal{D}-\mathcal{N}+1;1+\frac{1}{\bar{n}}\right)}\,.
\end{align}
For the case of $\delta n=\frac{\Lambda}{\phi +\Lambda}(n_{ex}-\bar{n})<0$, the ratio of the second and the third entries becomes $\frac{\mathcal{N}}{\mathcal{N}+\mathcal{D}}>1+\frac{1}{\bar{n}}$. Therefore, the ratio between the above two hypergeometric functions is bounded, and the fraction of the condensate becomes $ \frac{\langle n_0\rangle}{\langle N\rangle}\to 0$ in the large-$D$ limit. One can similarly show that the limit $\delta n\to 0$ exists and $\underset{\delta n\to 0}{\lim}\frac{\langle n_0\rangle}{\langle N\rangle}= 0$. 

On the other hand, notice that
\begin{align}
    \frac{_2F_1(a+1,-b+1;-c+1;d)}{_2F_1(a,-b;-c;d)}=\frac{\sum_{m}^{b-1} (m+1)_{(a)}\dfrac{(b-m)_{(m)}}{(c-1)_{(m)}}d^m }{\sum_{m}^b (m+1)_{(a-1)}\dfrac{(b-m+1)_{(m)}}{(c)_{(m)}}d^m }=\frac{\sum_m^b (m+a)(1-\frac{m}{b})(1+\frac{m}{c-1})f(m)}{\sum_m^b f(m)}\,,
    \label{sm:ratiosum}
\end{align}
where $f(m)$ is the distribution function of $m$ defined by 
\begin{align}
    f(m)=\frac{(m+1)_{(a-1)}}{b^{a-1}}\dfrac{(b-m+1)_{(m)}}{(c)_{(m)}}d^m\,.
\end{align}
For the case we consider here, from Eq.~\eqref{sm:n_critical} we know that the peak of the distribution is at 
\begin{align}
    m_{p}=\left(n_{ex}-\frac{(D-1)\Lambda-2\phi}{(D+1)\Lambda}\bar{n}\right)\frac{(D+1)\Lambda}{\phi+\Lambda}-\frac{\phi+\Lambda}{\chi}+1\,.
\end{align}
In the large-$D$ limit, 
\begin{align}
    m_{p}\sim \left(n_{ex}-\bar{n}\right)\frac{\Lambda}{\phi+\Lambda}D\,,
\end{align}
which is approximately proportional to $D$. Therefore, in the zeroth order approximation, 
\begin{align}
     \frac{ _2F_1(2,1-\mathcal{N};-\mathcal{D}-\mathcal{N}+2;1+\frac{1}{\bar{n}})}{ _2F_1\left(1,-\mathcal{N};-\mathcal{D}-\mathcal{N}+1;1+\frac{1}{\bar{n}}\right)}\approx m_p (1-\frac{m_p}{\mathcal{N}})(1+\frac{m_p}{\mathcal{N}+\mathcal{D}})= \delta n (1-\frac{\delta n}{\bar{n} +\delta n})(1+\frac{\delta n}{\bar{n}+\delta n+1})D\,,
\end{align}
where $\delta n=\frac{\Lambda}{\phi +\Lambda}(n_{ex}-\bar{n})$. The ratio of the two hypergeometric functions in Eq.~\eqref{sm:fra} scales linearly with $D$. Near the transition point, the fraction of the condensate given by Eq.~\eqref{sm:fra} is then approximated as:
\begin{align}
      \frac{\langle n_0\rangle}{\langle N\rangle} &\simeq  \frac{(1+\frac{1}{\bar{n}})}{(\bar{n}+\delta n+1)}  (1-\frac{\delta n}{\bar{n} +\delta n})(1+\frac{\delta n}{\bar{n}+\delta n+1})\delta n \nonumber\\
      & \simeq \frac{1}{\bar{n}}\dfrac{\Lambda}{\phi +\Lambda}(n_{ex}-\bar{n})\,.
\end{align}
Two important messages can be derived from the above calculation: the mean fraction of the condensation is nonzero in the $D\to \infty$ limit for $\left(n_{ex}-\bar{n}\right)>0$, which indicates the second-order phase transition in the large-$D$ limit, and that the critical exponent of the phase transition is $\beta=1$.
%Even if we do not constrain to the zeroth order, we can still see that:
%\begin{align}
%   \frac{\sum_{m}^{b-1} (m+1)_{(a)}\dfrac{(b-m)_{(m)}}{(c-1)_{(m)}}d^m }{\sum_{m}^b (m+1)_{(a-1)}\dfrac{(b-m+1)_{(m)}}{(c)_{(m)}}d^m }\simeq \frac{ b \int_0^\infty dx \ x(1-x)(1+l x)f(x)}{\int_0^\infty dx f(x)}\sim b \,,
%\end{align}

%\include{sm:phase_transition}
%\bibliography{bib}
\end{widetext}
\end{document}